\documentclass[amssymb,nofootinbib,nobibnotes,aps,prd,preprintnumbers]{revtex4}

 \usepackage{here}

\usepackage[dvipdfmx]{graphicx}
\usepackage{amssymb,amsfonts,amsmath,bm,color,multirow,cases,empheq,hyperref}
\usepackage{mathrsfs}

\usepackage{enumerate}
\usepackage{ulem}
\usepackage{afterpage}

\hypersetup{%
 setpagesize=false,
 bookmarksnumbered=true,%
 bookmarksopen=true,%
 colorlinks=true,%
 linkcolor=blue,%
 citecolor=red}

\begin{document}




\title{Innermost stable circular orbits around a spinning black hole binary}

\author{Eiki Kagohashi}
\email{sd19024@toyota-ti.ac.jp}
\author{Ryotaku Suzuki}
\email{sryotaku@toyota-ti.ac.jp}

\author{Shinya Tomizawa}
\email{tomizawa@toyota-ti.ac.jp}
\affiliation{Mathematical Physics Laboratory, Toyota Technological Institute\\
Hisakata 2-12-1, Nagoya 468-8511, Japan}
\date{\today}

\preprint{TTI-MATHPHYS-28}




\begin{abstract}

Using the exact solution that describes multi-centered rotating black holes, recently discovered by Teo and Wan, we investigate the innermost stable circular orbit (ISCO) for massive particles and the circular orbit for massless particles moving around a spinning black hole binary. We assume equal masses $M_1 = M_2=m$ and equal spin angular momenta $|J_1| = |J_2|$ for both black holes.
Firstly, we examine the case where two black holes are spinning in the same direction ($J_1=J_2$). 
We clarify that that for particles rotating in the same direction as  (opposite directions to)  black holes' spin, the greater the spin angular momenta of the black holes, the more the radii of the ISCO for massive particles and the circular orbit for massless particles decrease (increase). 
We show that distinct ISCO transitions occur for particles rotating in the same direction as the black holes in three ranges of spin angular momenta: $0<J_1/m^2=J_2/m^2< 0.395\ldots$, $0.395\ldots<J_1/m^2=J_2/m^2< 0.483\ldots$, and $0.483\ldots<J_1/m^2=J_2/m^2<0.5$. 
Conversely, particles rotating in the opposite direction to the black holes exhibit a consistent transition pattern for the case  $0<J_1/m^2=J_2/m^2<0.5$.
Secondly, we study the situation where binary black holes are spinning in opposite directions  ($J_1=-J_2$). 
We clarify that for large (small) separations between black holes, the ISCO appears near the black hole that is spinning in the same (opposite) direction as particles' rotation. 
Additionally, we show that different ISCO transitions occur in the three angular momentum ranges: $0<J_1/m^2=-J_2/m^2< 0.160\ldots$, $0.160\ldots<J_1/m^2=-J_2/m^2< 0.467\ldots$, and $0.467\ldots<J_1/m^2=-J_2/m^2<0.5$.

\end{abstract}

\date{\today}
\maketitle



\section{Introduction}
The dynamics of freely moving test particles in curved spacetimes, known as the geodesic structure, provide crucial insights into the gravitational field and geometry. In stationary spacetimes, there exist stationary orbits of particles, which follow geodesics along timelike Killing fields. Additionally, if the spacetime is also axisymmetric, these stationary orbits can take the form of circular orbits. Such fundamental orbits, associated with spacetime symmetries, are invaluable for understanding various observable phenomena (such as stellar motion and black hole shadows) around black holes. 
In the Schwarzschild black hole spacetime, there are both stable and unstable circular orbits for particles~\cite{Y.Hagihara:1931, K.Samuil :1949}. 
Let $r$ denote the circumference radius, and $M$ the mass of the black hole. We know that stable circular orbits exist in the range $r \geq 6GM/c^2$, and unstable circular orbits exist in the range $3GM/c^2 < r < 6GM/c^2$, where $G$ is the gravitational constant, and $c$ is the speed of light.
At the boundary $r=6GM/c^2$, there exists the innermost stable circular orbit (ISCO), and at $r = 3GM/c^2$, there exists the unstable photon circular orbit, which marks the last circular orbit. These phenomena are fundamental in the study of physical phenomena near a black hole. 
In particular, the ISCO is significant because it represents the closest stable circular orbit that a particle can have before inevitably falling into the black hole or the compact object due to gravitational attraction. 
Understanding the ISCO is crucial for studying accretion processes, orbital dynamics, and gravitational wave emission from compact object binaries. It provides insights into the behavior of matter and radiation in extreme gravitational environments near black holes and neutron stars.
For rotating black holes, such as the Kerr black hole, the ISCO location depends on the spin parameter of the black hole. Rotating black holes have a smaller ISCO radius compared to non-rotating ones, and the ISCO location moves closer to the event horizon as the spin of the black hole increases~\cite{Wilkins:1972rs,Bardeen:1972fi}.
Furthermore, such studies on circular orbits have been extended to charged non-rotating and rotating black holes as well~\cite{Pugliese:2010ps,Hackmann:2013pva,Cebeci:2015fie}. 
Recently, the related topics were discussed on non-singular black holes~\cite{Isomura:2023oqf,Garcia:2013zud,Stuchlik:2014qja,Zhou:2011aa,Eiroa:2010wm,Stuchlik:2019uvf,Gao:2020wjz,Rayimbaev:2020hjs,Carballo-Rubio:2022nuj,Novello:1999pg,Novello:2000km,Toshmatov:2019gxg,Kumar:2020ltt,Stuchlik2019,Toshmatov:2021fgm}.

\medskip
Exact solutions describing a black hole binary are of significant interest in astrophysics and theoretical physics because they provide insights into the behavior of black hole binary systems, such as those observed in gravitational wave astronomy. However, finding such solutions is a difficult problem due to a lack of symmetry and the time dependence of dynamics. Nevertheless, a class of static and stationary solutions describing a black hole binary has been known.
Initially, Israel and Khan~\cite{Israel1964}  discovered an exact solution that describes static multi-Schwarzschild black holes arranged along a rotational axis. However, the static equilibrium of these black holes necessitates the presence of conical singularities connecting them to counterbalance their gravitational attraction.
Furthermore, as a generalization of the double Schwarzschild black hole solution to a rotational case, Kramer and Neugebauer~\cite{Kramer1980} constructed the double Kerr solution, which describes the spacetime geometry around two rotating black holes. Although the gravitational fields of the two rotating black holes interact, leading to complex spacetime geometries, this solution also suffers from conical singularities between the two black holes.
Therefore, since vacuum exact solutions of a black hole binary are singular, it is more challenging to predict the motion of particles and light around a black hole binary, hindering the goal of future observations of their shadows.
However, there have been attempts~\cite{Wunsch:2013st,Dolan:2016bxj,Nakashi:2019mvs} to understand the characteristics of binary black holes by using the Majumdar-Papapetrou solution~\cite{Majumdar:1947eu,Papapetrou}, which is a static exact solution describing multi-black holes to the Einstein-Maxwell equations. 
For this solution, multiple black holes achieve static equilibrium by balancing the gravitational and electrostatic forces between two charged black holes, making the calculations straightforward. 

\medskip
Previous studies~\cite{Wunsch:2013st,Dolan:2016bxj,Nakashi:2019mvs} have analyzed the orbits of particles around the black holes, using the Majumdar-Papapetrou solution.
In particular, in Ref.~\cite{Nakashi:2019mvs}, Nahashi and Igata investigated the sequence of stable circular orbits for massive/massless particle moving around the Majumdar-Papapetrou dihole spacetime with equal mass, by reducing the particle motion to a two-dimensional potential problem.
In terms of qualitative differences of their sequences, they classified the separation between black holes into five ranges and found four critical values as the boundaries. 
When the separation is sufficiently large, the sequence of stable circular orbits has two branches, where one of two is on the symmetric plane of two black holes, whereas the other is on an arc-like curve connecting two black holes. 
In a certain separation range, the sequence on the symmetric plane separates into two parts, where one include infinity and the other  does not.  Moreover for a  smaller separation, the latter vanishes to become one.
They also clarified the dependence of the radii of marginally stable circular orbits and the ISCOs on the separation parameter, and found there is a discontinuous transition of the ISCO's radius, and additionally the separation range at which the radius of the ISCO can be smaller than that of the stable circular photon orbit.

 \medskip
 However, it is generally believed that each horizon of a realistic black hole binary is spinning.
Israel and Wilson~\cite{Israel:1972vx}, along with Perjes~\cite{Perjes:1971gv}, generalized the Majumdar-Papapetrou solution to a rotating solution in Einstein-Maxwell equation. 
However, it was later revealed that this solution represents a combination of naked singularities rather than black holes. 
It seems that achieving an equilibrium superposition of rotating black holes is not feasible within the framework of Einstein-Maxwell theory. 
Recently, Teo and Wan~\cite{Teo:2023wfd} succeeded in constructing a new regular exact solution that describes multi-centered spinning black holes in five-dimensional Kaluza-Klein theory. 
This solution, when dimensionally reduced, describes a balanced superposition of any number of dyonic rotating black holes in Einstein-Maxwell-dilaton theory. 
It includes parameters such as mass, spin angular momentum, and the position of each black hole. 
Moreover, when all spin angular momenta approach zero, the solution recovers the Majumdar-Papapetrou solution, as the scalar field also vanishes in this limit.

\medskip
Therefore, in this paper, as an extension of the study performed on the Majumdar-Papapetrou dihole spacetime~\cite{Nakashi:2019mvs}  to include spinning cases, we aim to investigate the circular orbits of particles around a spinning black hole binary  using the Teo-Wan solution.
In this study, we assume that the two black holes have equal mass and equal spin angular momenta (including spins in the same and opposite directions), and we vary each parameter to analyze the circular orbits of both massive particles and massless particles such as photons.
By incorporating the rotation of black holes into the calculations, our aim is to compute the particle orbits under conditions closer to reality than those of particle orbits in the Majumdar-Papapetrou solution.
In particular, we discuss the stability of circular orbits for massive and massless particles, and consider the effect of the spin of a black hole binary on the region of existence where the innermost stable circular orbits for massive particles exist, as well as the stability of photon orbits.

\medskip
In the following section~\ref{sec:TW}, 
we provide the minimal knowledge of the Kaluza-Klein multi-black hole solution necessary for our analysis.
In Sec.~\ref{sec:formalism}, we derive the conditions of stability for circular orbits of massive particles in the spinning black hole binary with equal mass and equal angular momenta. 
These conditions are extressed in terms of a two-dimensional effective potential.
In Sec.~\ref{sec:anaysis}, 
we study how the sequence of stable circular orbits depends on the separation between two spinning black holes and the spins of the black holes. 
In particular, for several typical values of spin angular momenta of black holes, 
we discuss the transition of the innermost stable circular orbit by the separation parameter. 
We devote Sec.~\ref{sec:summary}  to a summary and discussions.

\section{Review of Teo-Wan solutions}\label{sec:TW}

In this section, we review the Kaluza-Klein multi-black hole solution recently constructed by Teo and Wan~\cite{Teo:2023wfd}.
The Rasheed-Larsen solution describes the most general rotating black hole in five-dimensional Kaluza-Klein theory. The under-rotating extremal limit of this solution corresponds to a general class of solutions of Kaluza-Klein theory found by Clement~\cite{Clement:1986bt}, specified by two harmonic functions on a three-dimensional flat base space. 
Teo and Wan generalized this single extremal black hole solution to a multi-black hole solution by replacing the two harmonic functions, each having a single point source, with ones having an arbitrary number of point sources. In the dimensionally reduced four-dimensional spacetime, each black hole has its mass, spin angular momentum, electric and magnetic charges, which are set to be equal for simplicity, and each black hole is rotating with parallel or antiparallel spin vectors. This solution describes a physical spacetime such that there are neither naked singularities nor closed timelike curves both on and outside each black hole.

\medskip

Let us start with the five-dimensional Kaluza-Klein theory, in which the metric can be written as
\begin{eqnarray}
ds^2=e^{-\frac{2\phi}{\sqrt{3}}}(dx^5+A_\mu dx^\mu)^2+e^{\frac{\phi}{\sqrt{3}}}g_{\mu\nu} dx^\mu dx^\nu,
\end{eqnarray}
where the function $\phi$, the $1$-form $A_\mu$ and the four-dimensional metric $g_{\mu\nu}$  ($\mu,\nu=0,\ldots,3$)  do not depend on the fifth spacial coordinate $x^5$ which has a period of $2\pi R_{KK}$. 
As is well-known, the dimensional reduction of the five-dimensional Einstein theory leads the four-dimensional Einstein-Maxwell-dilaton theory, described by the action 
\begin{eqnarray}
S=\int d^4x\sqrt{-g} \left( R-\frac{1}{2}\partial_\mu \phi\partial^\mu \phi -\frac{1}{4}e^{-\sqrt{3} \phi} F^2 \right),
\end{eqnarray}
where $R$ is the Ricci scalar of the four-dimensional spacetime, $g={\rm det}(g_{\mu\nu})$, $F_{\mu\nu}=\partial_\mu A_\nu-\partial_\nu A_\mu$.
From this, the fields equations, the Einstein equation, the equations for the gauge potential $A_\mu$ and the scalar fields $\phi$ can be written as, respectively, 
\begin{eqnarray}
R_{\mu\nu}=\frac{1}{2}\partial_\mu \phi \partial_\nu \phi +\frac{1}{2}e^{-\sqrt{3} \phi} \left(F_{\mu\rho}F_\nu{}^\rho-\frac{1}{4} g_{\mu\nu} F^2 \right),
\end{eqnarray}
\begin{eqnarray}
\nabla_\mu\left(e^{-\sqrt{3} \phi}F^{\mu\nu}\right)=0,
\end{eqnarray}
\begin{eqnarray}
\nabla_\mu\nabla^\mu\phi=-\frac{\sqrt{3}}{4} e^{-\sqrt{3} \phi} F^2.
\end{eqnarray}

The four-dimensional metric, the gauge potential and the scalar field of the multi-centered rotating black hole solution, found by Teo and Wan~\cite{Teo:2023wfd}, can be written, respectively,  as
\begin{eqnarray}
ds^2_{(4)}&=&-(H_+H_-)^{-\frac{1}{2}}(dt+{\bm \omega^0}\cdot d{\bm x})^2+(H_+H_-)^{\frac{1}{2}}d{\bm x}\cdot d{\bm x},\\
{\bm A}&=&\frac{\sqrt{2}}{H_-}\left[-[(1+f)f-2g]dt+[(1+f){\bm\omega^0}+ H_- \tilde{\bm \omega}^5]\cdot d{\bm x} \right],\\
\phi&=&\frac{\sqrt{3}}{2}\ln \frac{H_+}{H_-},
\end{eqnarray}
where the functions $H_\pm$, one-forms ${\bm \omega}^0$, $\tilde{\bm \omega}^5$ on three-dimensional Euclid space ${\mathbb E}_3$ are given by
\begin{eqnarray}
H_\pm&=&(1+f)^2\pm 2g,\\
{\bm \omega}^0\cdot d{\bm x}&=&-\sum_{n=1}^N\frac{2J_n[(y-y_n)dx-(x-x_n)dy]}{|{\bm x}-{\bm x}_n|^3},\\
\tilde{\bm \omega}^5\cdot d{\bm x}&=&\sqrt{2}\sum_{n=1}^N\frac{M_n(z-z_n)[(y-y_n)dx-(x-x_n)dy]}{|{\bm x}-{\bm x}_n|[(x-x_n)^2+(y-y_n)^2]},
\end{eqnarray}
with the two harmonic functions, $f$ and $g$, having point sources at the positions ${\bm x}={\bm x_n}:=(x_n,y_n,z_n)$ ($n=1,\ldots,N$) on ${\mathbb E}^3$,
\begin{eqnarray}
f=\sum_{n=1}^N\frac{M_n}{|{\bm x}-{\bm x_n}|},\quad 
g=\sum_{n=1}^N\frac{J_n(z-z_n)}{|{\bm x}-{\bm x_n}|^3}.
\end{eqnarray}
This solution describes an asymptotically flat, stationary multi-centered rotating dyonic black holes, with each having an extremal horizon.
The $n$-th black hole at the position ${\bm x}={\bm x}_n$ on ${\mathbb E}^3$ carries the mass $M_n$, spin angular momentum $J_n$, equal electric and magnetic charges  $Q_n$ and $P_n$, respectively, given by $Q_n=P_n=M_n/\sqrt{2}$.  
Furthermore, the regularity of the metric on the horizon requires the  condition
\begin{eqnarray}
|J_n|<\frac{M_n^2}{2}, \label{eq:regularity}
\end{eqnarray}
since the horizon area, $4\pi \sqrt{M_n^4-4J_n^2}$, vanishes if this is saturated. 
Moreover, as shown in Ref.~\cite{Teo:2023wfd}, under the condition~(\ref{eq:regularity}), the spacetime is free from closed timelike curves on and outside the horizon. 
At the limit as $J_n\to 0$ for all $n$, the scalar field $\phi$ vanishes, and consequently, the Majumdar-Papapetrou solution describing static multiple dyonic black holes is restored.

\medskip
In the following section, we will delve into analyzing the stability of circular orbits for particles orbiting a spinning black hole binary. We specifically focus on the two-black-hole solution (denoted as $N=2$) expressed in cylindrical coordinates $(\rho, z)$, where these coordinates are defined by $(x, y) = (\rho \cos \varphi, \rho \sin \varphi)$. 
The solution is given by
\begin{eqnarray}
ds^2_{(4)}&=&-(H_+H_-)^{-\frac{1}{2}}(dt+\omega^0_\varphi d\varphi)^2+(H_+H_-)^{\frac{1}{2}}[d\rho^2+dz^2+\rho^2d\varphi^2],\\
{\bm A}&=&\frac{\sqrt{2}}{H_-}\left[-[(1+f)f-2g]dt+[(1+f){\omega_\varphi^0}+ H_- \tilde{ \omega_\varphi}^5]d\varphi \right],\\
\phi&=&\frac{\sqrt{3}}{2}\ln \frac{H_+}{H_-},
\end{eqnarray}
where 
\begin{eqnarray}
H_\pm&=&(1+f)^2\pm 2g,\\
\omega_\varphi^0&=&\frac{2J_1 \rho^2}{\sqrt{\rho^2+(z-a)^2}^3}+\frac{2J_2 \rho^2}{\sqrt{\rho^2+(z+a)^2}^3},\\
\tilde{\omega}_\varphi^5&=&-\sqrt{2}\frac{M_1(z-a)}{\sqrt{\rho^2+(z-a)^2}}-\sqrt{2}\frac{M_2(z+a)}{\sqrt{\rho^2+(z+a)^2}},
\end{eqnarray}
with two harmonic functions having point sources at positions ${\bm x}_1=(0,0,-a)$ and ${\bm x}_2=(0,0,a)$:
\begin{eqnarray}
f&=&\frac{M_1}{\sqrt{\rho^2+(z+a)^2}}+\frac{M_2}{\sqrt{\rho^2+(z-a)^2}},\\
g&=&\frac{J_1(z+a)}{\sqrt{\rho^2+(z+a)^2}^3}+\frac{J_2(z-a)}{\sqrt{\rho^2+(z-a)^2}^3}.
\end{eqnarray}
Here, two black hole are positioned along the $z$-axis, resulting in the spacetime possessing both an axial Killing vector $\partial/\partial\phi$ as well as a timelike Killing vector  $\partial/\partial t$. 
In the next section, we define the functions $A$ and $B$ as
\begin{eqnarray}
A(\rho,z)&:=&\omega_\varphi^0= \frac{2\rho^2J_1}{[\rho^2+ (z-a)^2]^{3/2}} + \frac{2\rho^2J_2}{[\rho^2+ (z-a)^2]^{3/2}},\\
B(\rho,z)&:=&\sqrt{H_+H_-}=[(1+f)^4-4g^2]^{1/2}. \label{defB}
\end{eqnarray}

\section{Our formalism}\label{sec:formalism}

Following the procedure outlined in Ref.~\cite{Nakashi:2019mvs}, we derive the conditions of stability for circular orbits of a massive particle moving around the dihole spacetime with equal mass and equal spin angular momenta. 
This particle motion can be reduced to motion in a two-dimensional effective potential $U(\rho,z)$ on the $(\rho,z)$-plane.
According to the linear stability analysis of circular orbits, a circular orbit is stable if and only if it exists at a local minimum point of $U$. 
We refer to such a circular orbit as a stable circular orbit. 
Conversely, a circular orbit is unstable if and only if it exists at a local maximum point of $U$ or a saddle point of $U$. 
We refer to such a circular orbit as an unstable circular orbit.
A stable circular orbit lies somewhere at stationary points of $U$, where $d U=0$.
Hence, the condition of a stable circular orbit, which exists at a local minimum point of $U$, is determined by the positivity condition for two eigenvalues of its Hesse matrix  at the stationary point.

\medskip
The four-velocity $u^\mu=\dot x^\mu:=dx^\mu/d\lambda$ of particles must satisfy the constraint
\begin{eqnarray}
 g_{\mu \nu} u^\mu u^\nu=    -\frac{1}{B}\dot t^2 - \frac{2A}{B}\dot t \dot \phi + \left(B\rho^2 - \frac{A^2}{B}\right)\dot \phi^2 + B(\dot\rho^2 + \dot z^2) \label{metricsAB} = -\kappa:=
     \begin{cases}
1 & {\rm (massive \ particles)} \\
0 &  {\rm (massless \ particles)}
\end{cases},\label{eq:4vel}
\end{eqnarray}
where we choose $ \lambda $ to represent the proper time for massive particles and the affine parameter for massless particles. 
From the Lagrangian for free massive or massless particles, 
\begin{align}
    \mathscr{L}=\frac{1}{2}g_{\mu\nu}\dot x^\mu \dot x^\nu,
\end{align}
the Euler-Lagrange equation can be expressed as
\begin{align}
 \frac{d}{d \lambda} \frac{\partial}{\partial \dot x^\alpha} g_{\mu \nu}\dot x^\mu \dot x^\nu -    \frac{\partial}{\partial x^\alpha}g_{\mu\nu}\dot  x^\mu \dot x^\nu  = 0. \label{eq:ELeq}
\end{align}
Since the time coordinate $t$ and the angular coordinate $\phi$ included in the Lagrangian are cyclic coordinates, 
the $t$ and $\phi$ components in Eq.~(\ref{eq:ELeq}) yield the constants of motion as
\begin{align}
   \frac{\partial}{\partial \dot t} g_{\mu\nu}\dot x^\mu \dot x^\nu =-2\varepsilon \ (={\rm const.}) &\Longleftrightarrow  g_{t\phi}\dot \phi + g_{tt}\dot t =-\varepsilon\\
  & \Longleftrightarrow
    \frac{A}{B}\dot \phi + \frac{1}{B}\dot t = \varepsilon, \label{defer}\\
    \frac{\partial}{\partial \dot \phi} g_{\mu\nu}\dot x^\mu \dot x^\nu =2j\ (={\rm const.}) &\Longleftrightarrow  g_{t\phi}\dot t+ g_{\phi \phi}\dot \phi=j \\
    &\Longleftrightarrow
    -\frac{A}{B}\dot t +\left(B\rho^2 - \frac{A^2}{B} \right)\dot \phi= j,
        \label{defjr}
\end{align}
where the constants $\varepsilon$ and $j$ represent the energy and angular momentum, respectively, for a particle with unit rest mass. 
By substituting Eqs.~(\ref{defer}) and (\ref{defjr}) into Eq.~(\ref{eq:4vel}) and  eliminating $\dot t$ and $\dot \phi$, we obtain
\begin{align}
    B(\dot \rho^2+\dot z^2)+ \frac{j^2}{B\rho^2} + \frac{2A}{B\rho^2}j\varepsilon-B\varepsilon^2 = -\kappa
  \Longleftrightarrow     B(\dot \rho^2+\dot z^2) + \varepsilon^2  U(\rho,z;\bar j) = -\kappa, \label{Vspin}
\end{align}
where 
\begin{align}
    U(\rho,z;j) 
    &= X(\rho,z)j^2 + Y(\rho,z) j + Z(\rho,z),
\end{align}
with
\begin{align}
    X(\rho,z):=\frac{1}{B\rho^2} ,\; Y(\rho,z) := \frac{2A}{B\rho^2},\;Z(\rho,z) := \frac{A^2}{B\rho^2}-B.
\end{align}
Here, $\bar j$ represents the angular momentum normalized by $\varepsilon$,  given by $\bar j:=j/\varepsilon$. 
Thus, the motion of particles around spinning black hole binary can be reduced to motion in the two-dimensional potential $U(\rho,z;\bar j)$.

\medskip
Let us consider the circular orbits of particles rotating around the $z$ axis. It follows from $\dot{\rho} = \dot{z} = 0$ and Eq.~(\ref{Vspin}) that the potential $U$ must satisfy
\begin{align}
       U(\rho,z;\bar j) =- \frac{\kappa}{\varepsilon^2}. \label{eq:U}
\end{align}
Moreover, since such particles lie at stationary points of the potential,  $U$ must satisfy
\begin{align}
      &U_{,\rho}(\rho,z;\bar j) =X_{,\rho}(\rho,z)\bar j^2+Y_{,\rho}(\rho,z)\bar j+Z_{,\rho}(\rho,z)= 0,  \label{Vrho0}\\ 
    &U_{,z}(\rho,z;\bar j) =X_{,z}(\rho,z)\bar j^2+Y_{,z}(\rho,z)\bar j+Z_{,z}(\rho,z)= 0. \label{Vz0}
\end{align}
From Eq.(\ref{Vrho0}), we can obtain
\begin{align}
    \bar{j}=\bar{j}_{0\pm}(\rho,z) := \frac{-Y_{,\rho} \mp \sqrt{Y_{,\rho}^2- 4X_{,\rho} Z_{,\rho}}}{2X_{,\rho}}, \label{defj0}
\end{align}
where we note $\bar{j}_{0+}(\rho,z)>0$ and $\bar{j}_{0-}(\rho,z)<0$, and  for simplicity, in the following, we represent either $\bar{j}_{0+}(\rho,z)$ or $\bar{j}_{0-}(\rho,z)$ as $\bar{j}_{0}(\rho,z)$.
Therefore, substituting this into Eq.~(\ref{Vz0}), we can represent the sequence  of circular orbits in the $(\rho,z)$-plane as follows:
\begin{align}
  S:=\{(\rho,z)|  U_{,z}(\rho,z;\Bar{j_0})= 0\}, \label{orbit1}
\end{align}
which denote certain curves in the two-dimensional $(\rho,z)$-plane. 
Additionally, from Eq.~(\ref{Vspin}) and the positivity of the function $B$, the allowed region of the particle motion is given by $U(\rho,z;\bar j_0)\le 0$, i.e.,  the two-dimensional region in the $(\rho,z)$-plane, 
\begin{eqnarray}
V:=\{(\rho,z)|U(\rho,z;\bar j_0)>0 \},
\end{eqnarray}
corresponds to the forbidden region of the particle motion.

\medskip
Finally, we further impose the stability conditions on circular orbits.
Let $H=(H_{ij}):=(\partial_j\partial_i U)$  $(i,j=\rho,z)$ be the Hesse matrix of $U$ on the two-dimensional flat space $d\rho^2+dz^2$.
Particles moving along stable circular orbits exist at local minima of $U$, and hence the stable circular orbits must be included in the  two-dimensional region $H$ in the $(\rho,z)$-plane defined by
\begin{eqnarray}
H:=\{(\rho,z)|  d_0(\rho,z)> 0, t_0(\rho,z)>0\},
\end{eqnarray}
where 
\begin{align}
  &d_0(\rho,z):=  {\rm det} H_{ij}(\rho,z;\Bar{j_0})> 0 \label{orbit2},\\
   &t_0(\rho,z):= {\rm tr} H_{ij}(\rho,z;\Bar{j_0}) > 0 \label{orbit22}.
\end{align}
Thus, we can visualize the sequence of stable circular orbits for massive particles by the curve $S$ inside the stable region $H$ and outside the forbidden region $V$ in the $(\rho,z)$-plane. 
Moreover, one observes from Eq.~(\ref{eq:U}) that the circular orbits for massless particles appear at the intersection of the curve $S$ and the boundary of the region $V$.

\section{ISCOs}\label{sec:anaysis}

We investigate circular orbits around the Teo-Wan black hole binary for various separation parameters $a$ between the two black holes along the $z$-axis, as well as the spin angular momenta of the black holes $(J_1, J_2)$. Specifically, we determine the positions of the ISCO and massless particles' circular orbits  for each parameter. 
 For simplicity, we focus on a binary system with equal masses $M_1 = M_2$ and equal magnitudes of spin angular momenta  $|J_1| = |J_2|$, assuming that the two black holes are spinning either in the same direction ($J_1 = J_2$) or in opposite directions ($J_1 = -J_2$). Hereafter, we use the unit of $M_1=M_2=1$.

\subsection{Two black holes spinning in the same direction}\label{subsec:samespin}

\begin{figure}[h]
     \begin{minipage}[t]{0.3\columnwidth}
    \centering
    \includegraphics[width = 4cm]{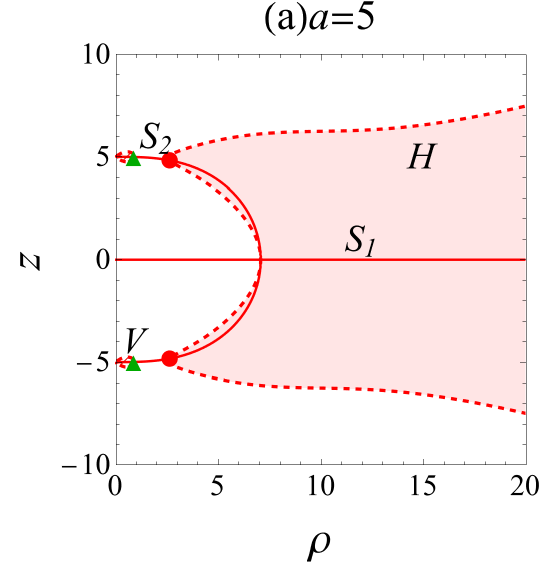}
    \includegraphics[width = 4cm]{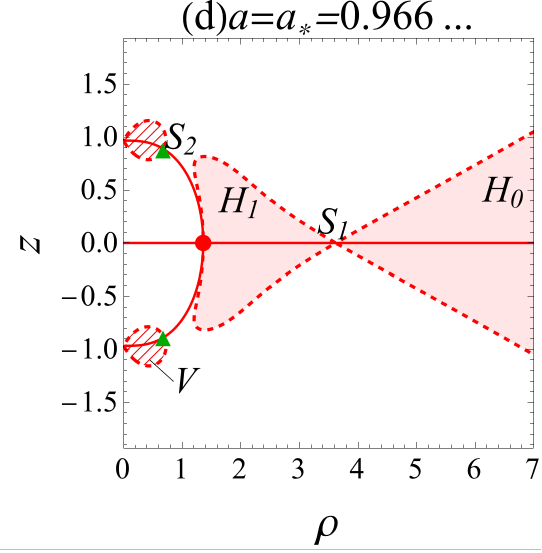}
    \includegraphics[width = 4cm]{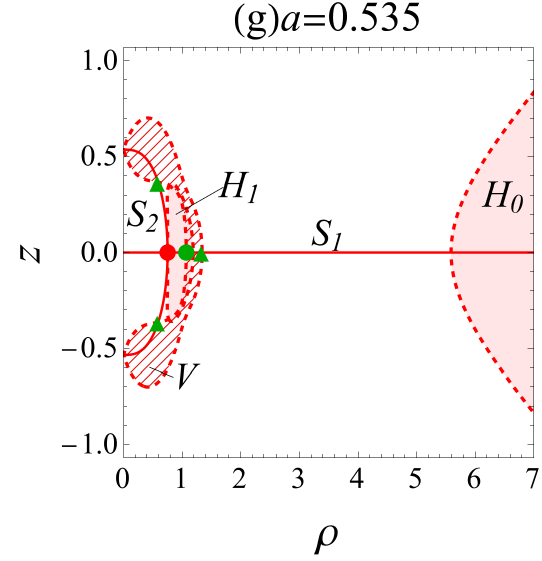}
    \end{minipage}
  \begin{minipage}[t]{0.3\columnwidth}
    \centering
    \includegraphics[width = 4cm]{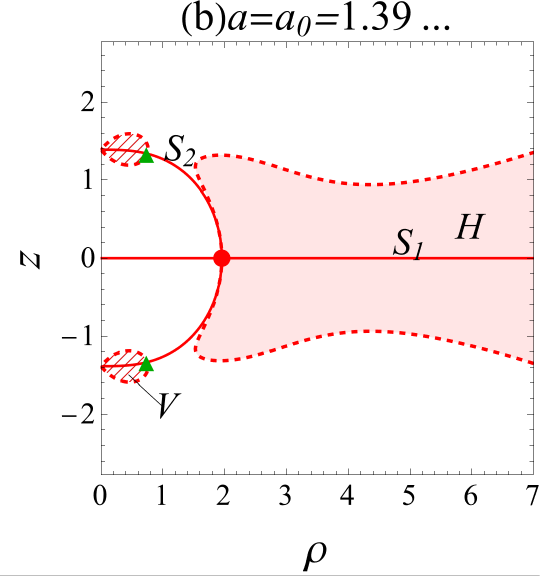}
    \includegraphics[width = 4cm]{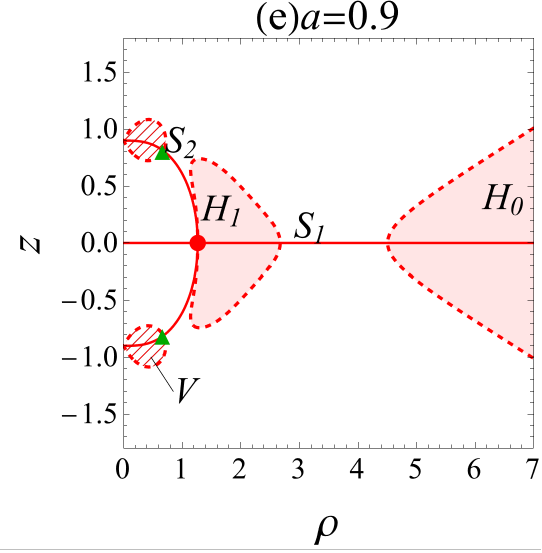}
\includegraphics[width = 4cm]{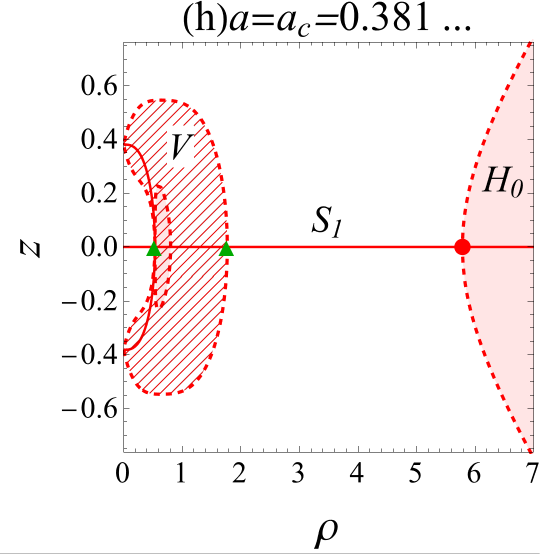}
  \end{minipage}
   \begin{minipage}[t]{0.3\columnwidth}
    \centering
    \includegraphics[width = 4cm]{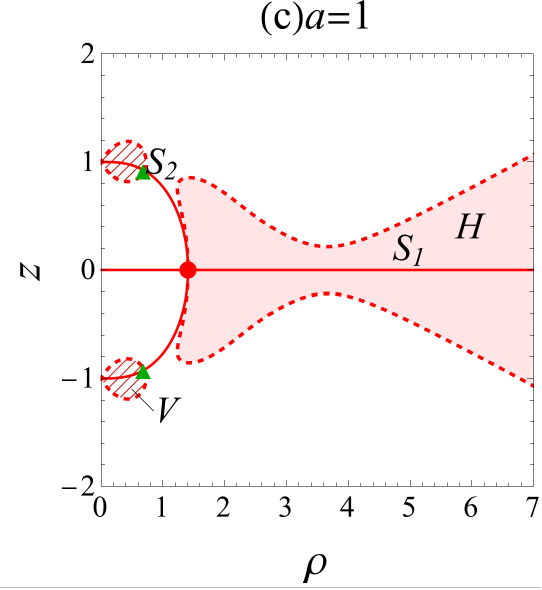}
    \includegraphics[width = 4cm]{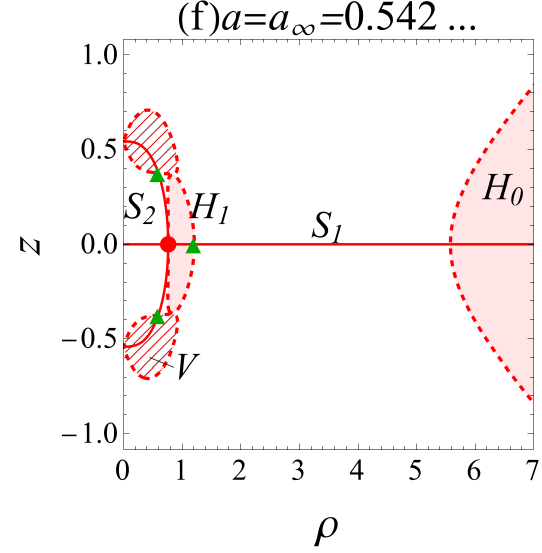}
    \includegraphics[width = 4cm]{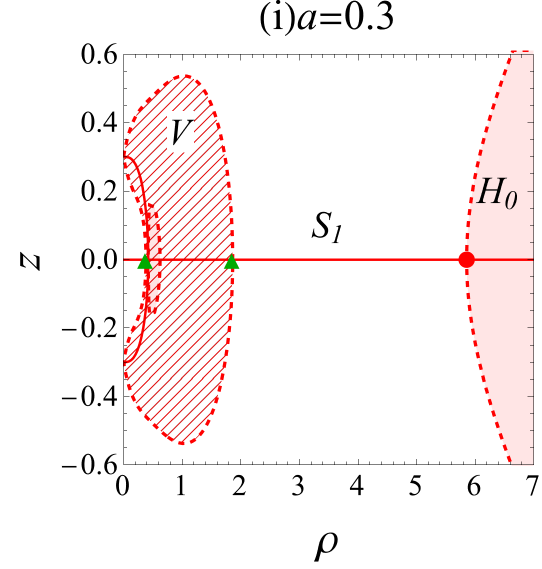}
   \end{minipage}
   
  \caption{Circular orbits for massive and massless particles with $j>0$ and $J_1=J_2=0.01$. 
Each panel differs in the separation $2a$ between two black holes located at $(\rho,z)=(0,\pm a)$. 
The sequence of stationary points $S$  comprises two curves, $S_1$ and $S_2$, represented by solid lines. 
Here, $S_1$ corresponds to the line $z=0$, while $S_2$ represents the arc connecting the two black holes. 
The red circle points represent ISCOs, while the green triangle and circle points denote unstable and stable circular orbits, respectively. 
Additionally, the red-colored region denotes the stability region $H$ and the hatched region denote the forbidden region $V$.
\label{j1+j20.01+}}
\end{figure}

First, we consider the case where two black holes have the same spin angular momenta, varying the value of $J_1=J_2$ from $0$ to nearly $0.5$. Due to the system's symmetry, the appearance of orbits is symmetric with respect to the $z=0$ plane.
In the following analysis, we illustrate the localization of circular orbits in the $(\rho,z)$-plane, as shown in Fig.~\ref{j1+j20.01+}. Here, the sequence of stationary points $S$ comprises two curves, $S_1$ and $S_2$, depicted by solid lines. $S_1$ corresponds to the line $z=0$, while $S_2$ represents the arc connecting the two black holes.
The red circle points on the plot represent the innermost stable circular orbits (ISCOs) for massive particles, while the green triangle and circle points denote unstable and stable circular orbits for massless particles, respectively. Additionally, the red-colored region denotes the stable region $H$, and the hatched region represents the forbidden region $V$.
Additionally, the red-colored region denotes the stable region $H$ and the hatched region denote the forbidden region $V$.
$S_2$ and the boundary of $H$ are always tangent at a point on $z=0$ for any $a$.

\subsubsection{Particles with a positive orbital angular momentum $(j>0)$}

\paragraph{$J_1=J_2=0.01$}

As shown in Fig.~\ref{j1+j20.01+}(a), for a sufficiently large separation ($a \gg 1$), stable circular orbits exist along $S_1$. Additionally, they also exist on the portion of $S_2$ extending from the intersection of $S_1$ and $S_2$ to the ISCOs near the two black holes. The unstable circular orbits for massless particles appear near the two black holes at the intersections of $S_2$ and the boundaries of $V$.
As $a$ decreases to $a_0 = 1.39...$, the two ISCOs approach the intersection of $S_1$ and $S_2$. At $a = a_0$, as depicted in Fig.~\ref{j1+j20.01+}(b), $S_2$ and the boundary of $H$ intersect only at $z=0$, resulting in the sequence of stable circular orbits existing only on $S_1$.
In Figs.~\ref{j1+j20.01+}(d) and (e), for $a \leq a_* = 0.966...$, $H$ is divided into two regions, $H_0$ (including infinity) and $H_1$ (excluding infinity). The ISCO appears at the intersection of $S_1$, $S_2$, and the boundary of $H_1$.
At $a = a_\infty = 0.542...$, as depicted in Fig.~\ref{j1+j20.01+}(f), another unstable circular orbit for massless particles appears at the intersection of $S_1$ and the boundary of $H_1$.
In Fig.~\ref{j1+j20.01+}(g), for $a_c = 0.381... < a < a_\infty$, a stable circular orbit exists for massless particles at the intersection of $S_1$ and the boundary of $V$ inside $H_1$, along with three unstable circular orbits.
In Figs.~\ref{j1+j20.01+}(h) and (i), for $a \leq a_c$, since the entire curve of $S_2$ is included in $V$, the stable circular orbits for massive particles appear only in $H_0$, where the ISCO appears at the intersection of $S_1$ and the boundary of $H_0$. Meanwhile, for massless particles, the stable circular orbit disappears, and two unstable circular orbits on $S_2$ merge into a single orbit on $z=0$.

\medskip
\paragraph{$J_1=J_2=0.4$}

\begin{figure}[h]

     \begin{minipage}[t]{0.3\columnwidth}
    \centering
      \includegraphics[width = 4cm]{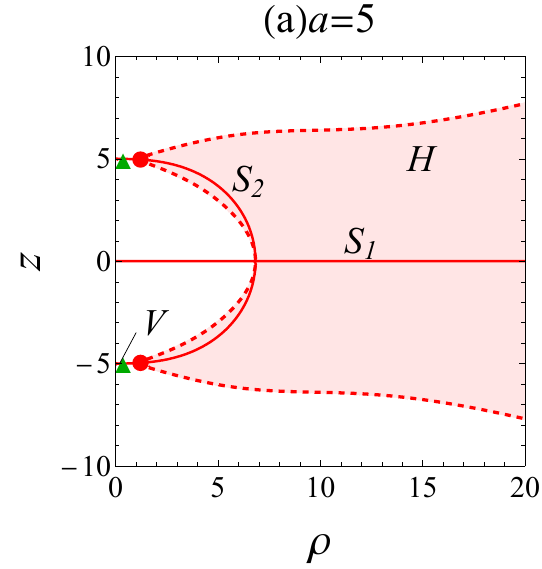}
   \includegraphics[width = 4cm]{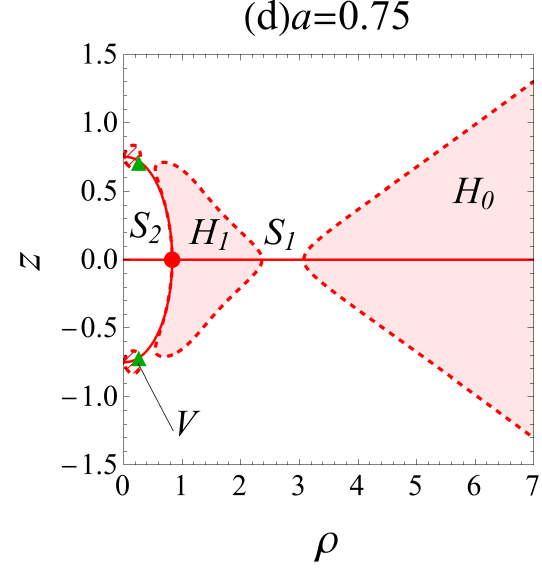}
   \includegraphics[width = 4cm]{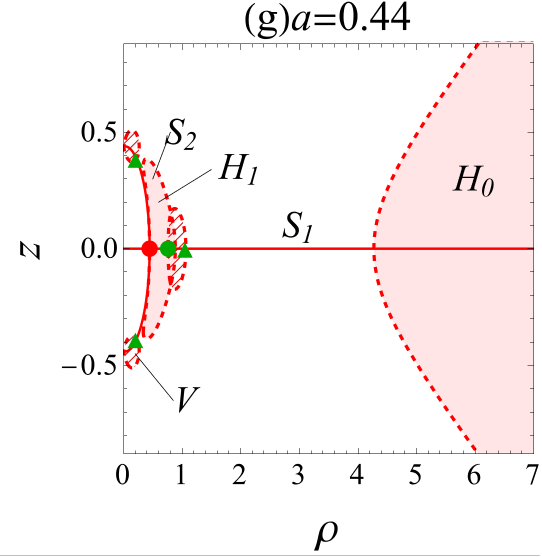}
    \end{minipage}
  \begin{minipage}[t]{0.3\columnwidth}
    \centering
       \includegraphics[width = 4cm]{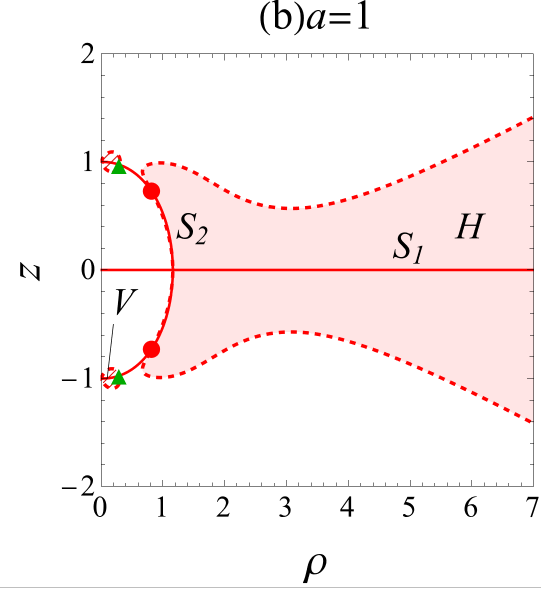}
 \includegraphics[width = 4cm]{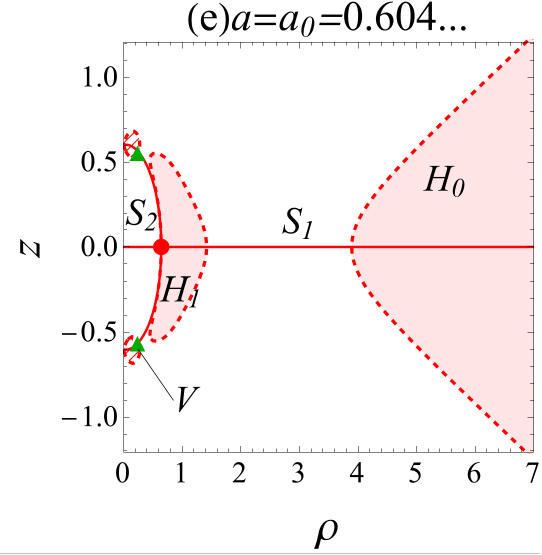}
   \includegraphics[width = 4cm]{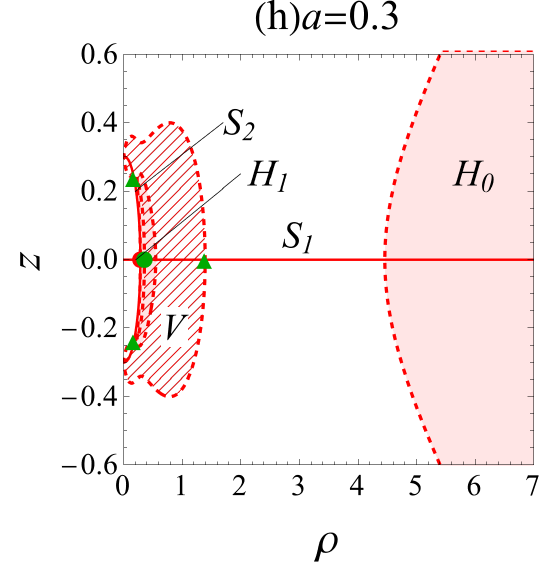}
  \end{minipage}
   \begin{minipage}[t]{0.3\columnwidth}
    \centering
       \includegraphics[width = 4cm]{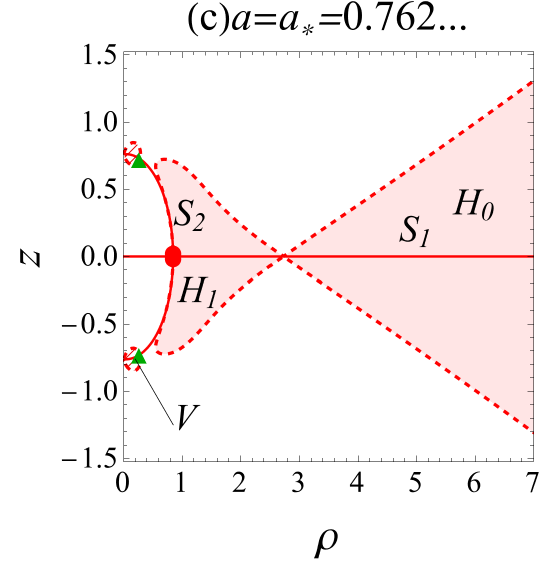}
   \includegraphics[width = 4cm]{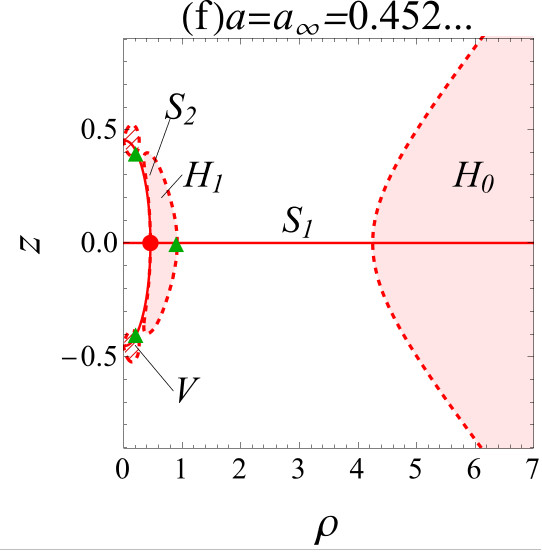}
   \includegraphics[width = 4cm]{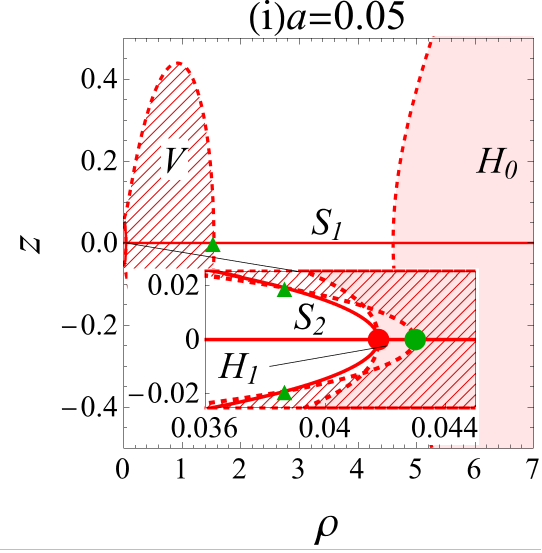}
 \end{minipage}

\caption{Circular orbits for massive and massless particles with $j>0$ and $J_1=J_2=0.4$. 
Each panel differs in the separation $2a$ between two black holes located at $(\rho,z)=(0,\pm a)$. 
The sequence of stationary points $S$  comprises two curves, $S_1$ and $S_2$, represented by solid lines. 
Here, $S_1$ corresponds to the line $z=0$, while $S_2$ represents the arc connecting the two black holes. 
The red circle points represent ISCOs, while the green triangle and circle points denote unstable and stable circular orbits, respectively. 
Additionally, the red-colored region denotes the stability region $H$ and the hatched region denote the forbidden region $V$. In the panel (i), we also show the closeup near $(\rho,z)=(0.04,0)$. 
\label{j1+j204+}}
\end{figure}

In this case,, circular orbits exhibit similar characteristics to the $J_1=J_2=0.01$ case. However, as evident from Fig.~\ref{j1+j20.01+}(a) and Fig.~\ref{j1+j204+}(a), the ISCOs approach the black holes due to the frame dragging effect.
As shown in Figs.~\ref{j1+j204+}(c) and (e), the transitions at $a=a_0$ and $a=a_*$ in the previous case now occur in reverse order. 
Furthermore, as indicated in Fig.~\ref{j1+j204+}(i), the transitions at $a=a_c$ in the $J_1=J_2=0.01$ case do not occur. Hence, for sufficiently large spin angular momenta, the ISCO remains at the intersection of $S_1$, $S_2$, and the boundary $H_1$. For massless particles, there exist three unstable circular orbits and a stable circular orbit for arbitrarily small $a$.

\medskip

\medskip
\paragraph{$J_1=J_2=0.499$}

\begin{figure}[h]

     \begin{minipage}[t]{0.3\columnwidth}
    \centering
      \includegraphics[width = 4cm]{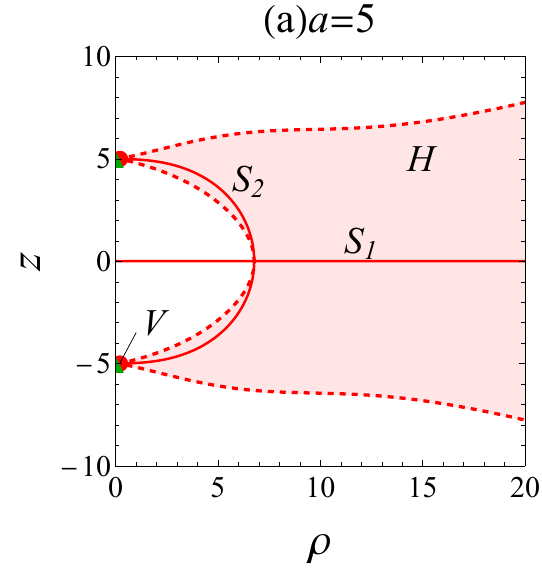}
   \includegraphics[width = 4cm]{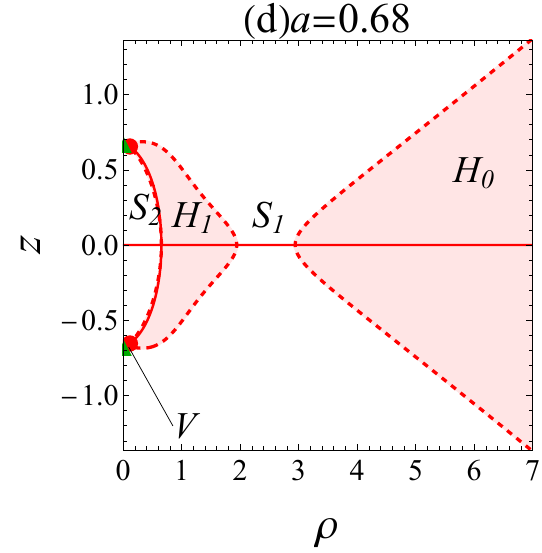}
   \includegraphics[width = 4cm]{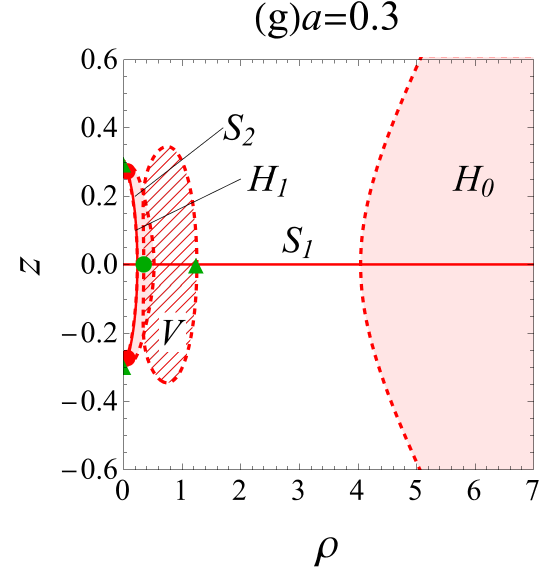}
    \end{minipage}
  \begin{minipage}[t]{0.3\columnwidth}
    \centering
       \includegraphics[width = 4cm]{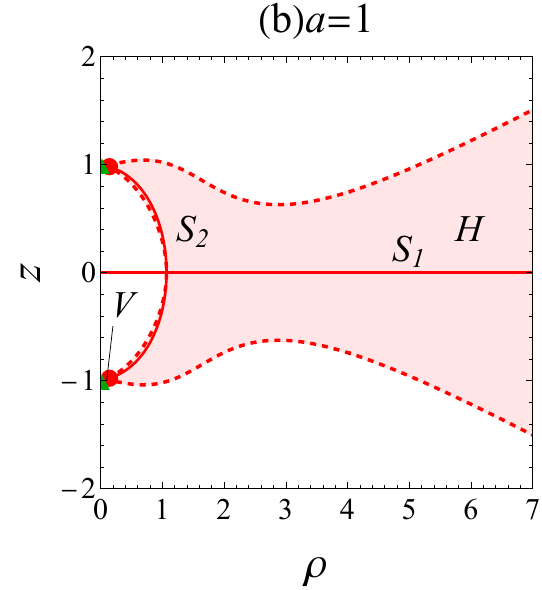}
 \includegraphics[width = 4cm]{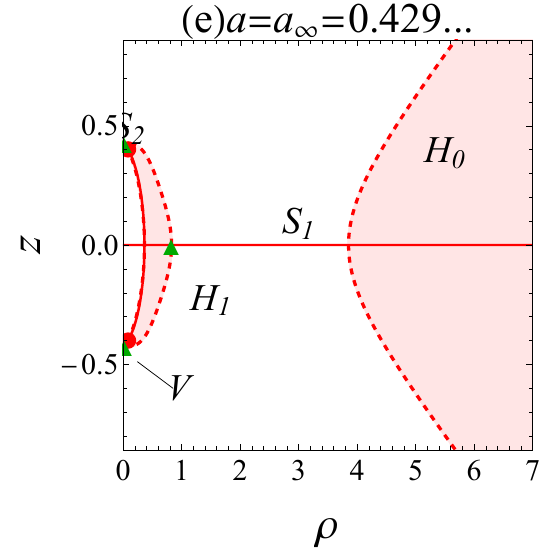}
   \includegraphics[width = 4cm]{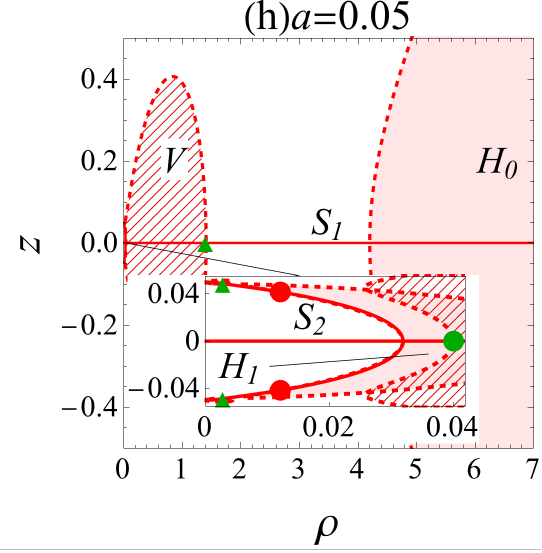}
  \end{minipage}
   \begin{minipage}[t]{0.3\columnwidth}
    \centering
       \includegraphics[width = 4cm]{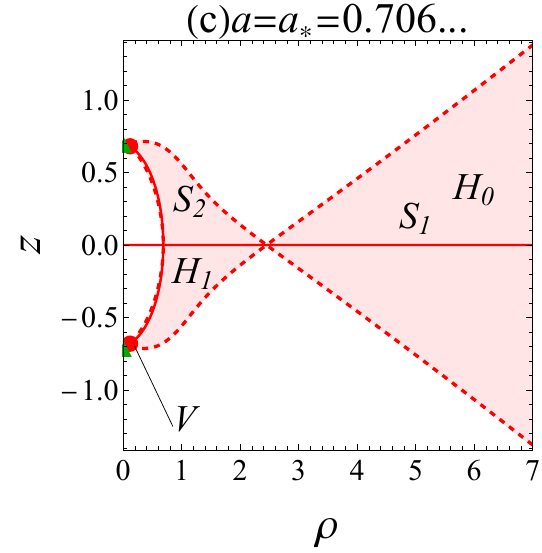}
   \includegraphics[width = 4cm]{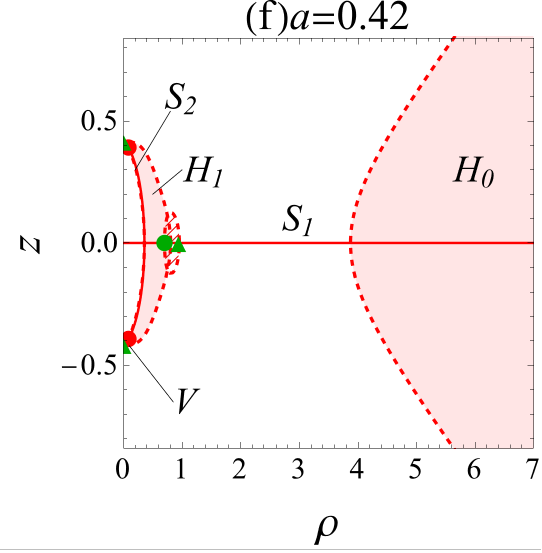}
 \end{minipage}

\caption{Circular orbits for massive and massless particles with $j>0$ and $J_1=J_2=0.499$. 
Each panel differs in the separation $2a$ between two black holes located at $(\rho,z)=(0,\pm a)$. 
The sequence of stationary points $S$  comprises two curves, $S_1$ and $S_2$, represented by solid lines. 
Here, $S_1$ corresponds to the line $z=0$, while $S_2$ represents the arc connecting the two black holes. 
The red circle points represent ISCOs, while the green triangle and circle points denote unstable and stable circular orbits, respectively. 
Additionally, the red-colored region denotes the stability region $H$ and the hatched region denote the forbidden region $V$.
In the panel (h), we also show the closeup near $(\rho,z)=(0.02,0)$.
\label{j1+j20499+}}
\end{figure}

As depicted in Fig.~\ref{j1+j20499+}(a), the ISCOs continue to approach the black holes. 
In this case, not only does the transition at $a=a_c$ cease to occur, but also the transition at $a=a_0$ appearing in the previous two cases. There are always two ISCOs at the intersections of $S_2$ and the boundary of $H_1$ for $a>0$, as shown in Fig.~\ref{j1+j204+}(h).

\subsubsection{Particles with a negative orbital angular momentum $(j<0)$}

\begin{figure}[h]

     \begin{minipage}[t]{0.3\columnwidth}
    \centering
      \includegraphics[width = 4cm]{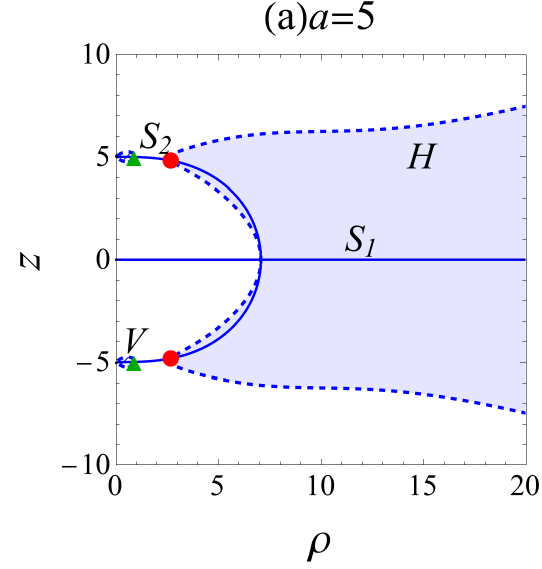}
   \includegraphics[width = 4cm]{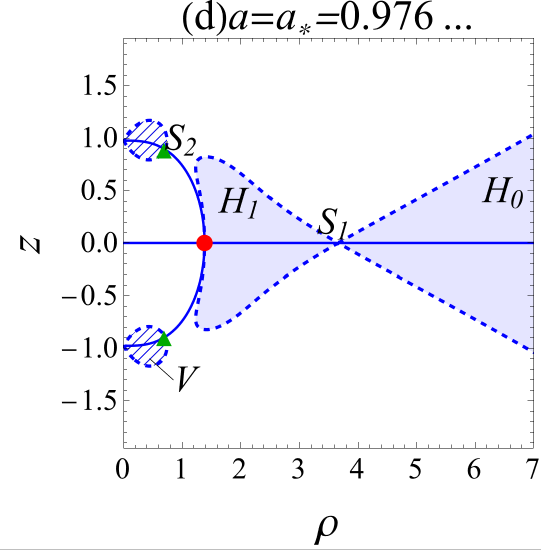}
   \includegraphics[width = 4cm]{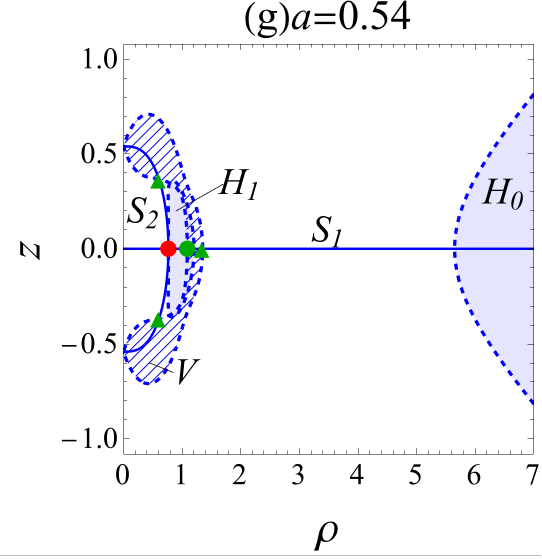}
    \end{minipage}
  \begin{minipage}[t]{0.3\columnwidth}
    \centering
       \includegraphics[width = 4cm]{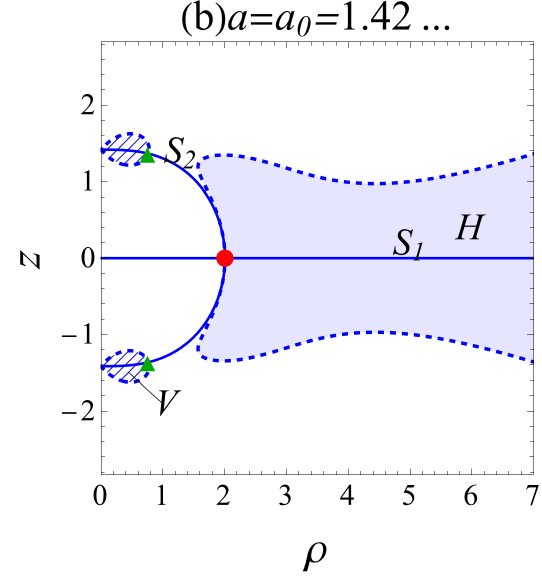}
 \includegraphics[width = 4cm]{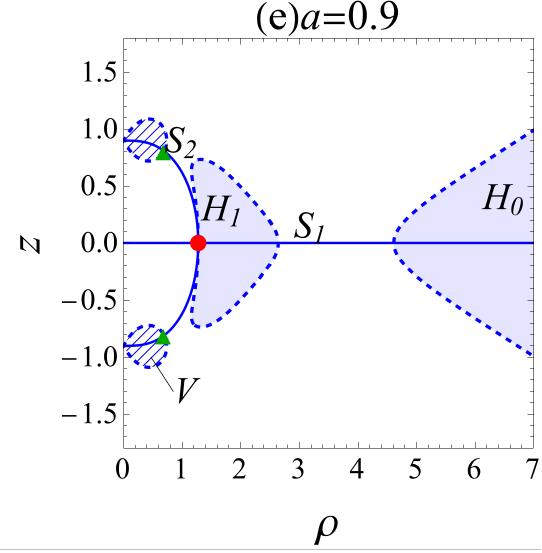}
   \includegraphics[width = 4cm]{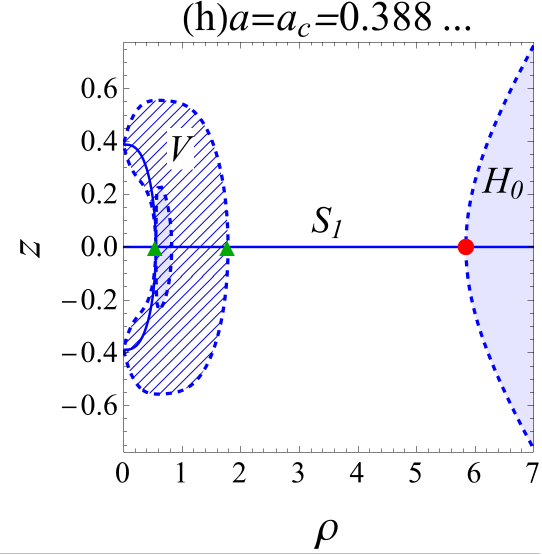}
  \end{minipage}
   \begin{minipage}[t]{0.3\columnwidth}
    \centering
       \includegraphics[width = 4cm]{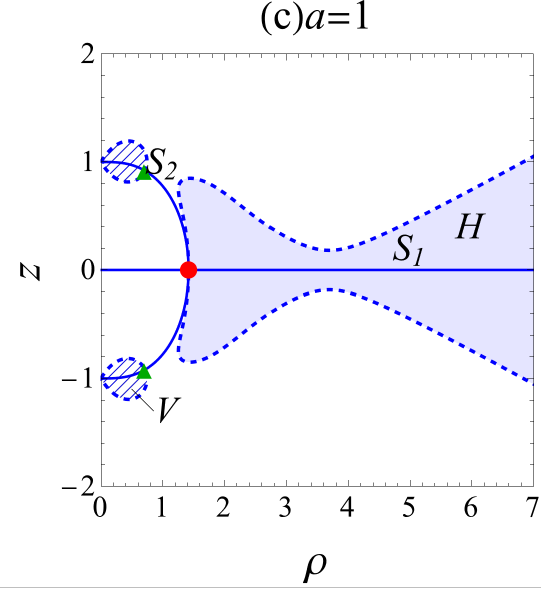}
   \includegraphics[width = 4cm]{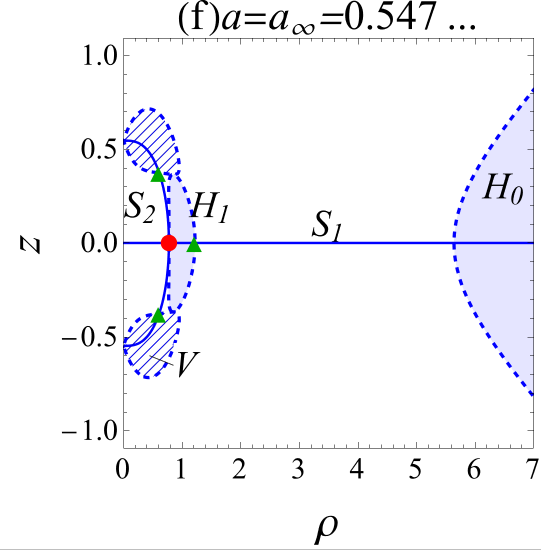}
   \includegraphics[width = 4cm]{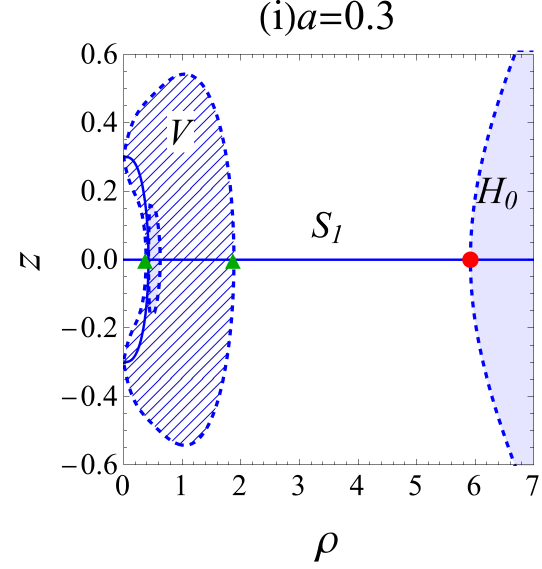}
 \end{minipage}
 
   \caption {Circular orbits for massive and massless particles with $j<0$ and $J_1=J_2=0.01$. 
Each panel differs in the separation $2a$ between two black holes located at $(\rho,z)=(0,\pm a)$. 
The sequence of stationary points $S$  comprises two curves, $S_1$ and $S_2$, represented by solid lines. 
Here, $S_1$ corresponds to the line $z=0$, while $S_2$ represents the arc connecting the two black holes. 
The red circle points represent ISCOs, while the green triangle and circle points denote unstable and stable circular orbits, respectively. 
Additionally, the red-colored region denotes the stability region $H$ and the hatched region denote the forbidden region $V$.
\label{j1+j20.01-}}

\end{figure}

\begin{figure}[h]

     \begin{minipage}[t]{0.3\columnwidth}
    \centering
      \includegraphics[width = 4cm]{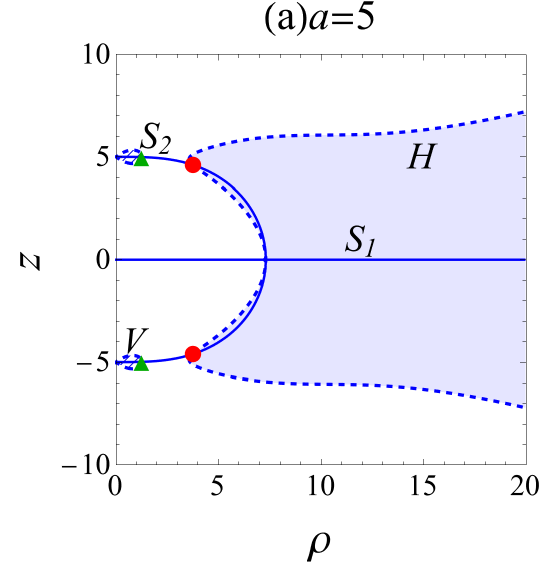}
   \includegraphics[width = 4cm]{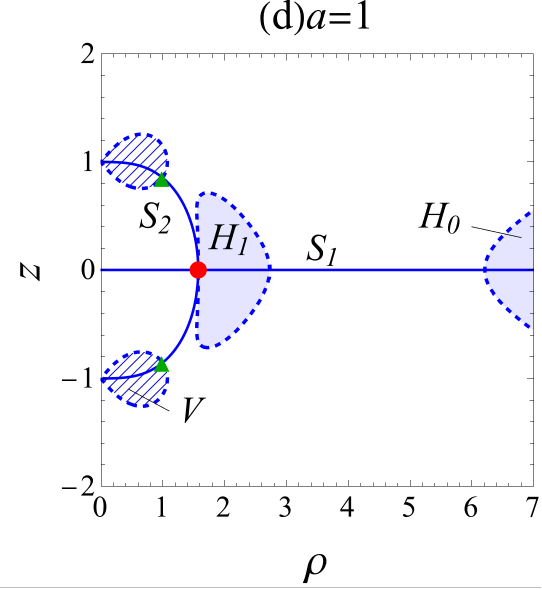}
   \includegraphics[width = 4cm]{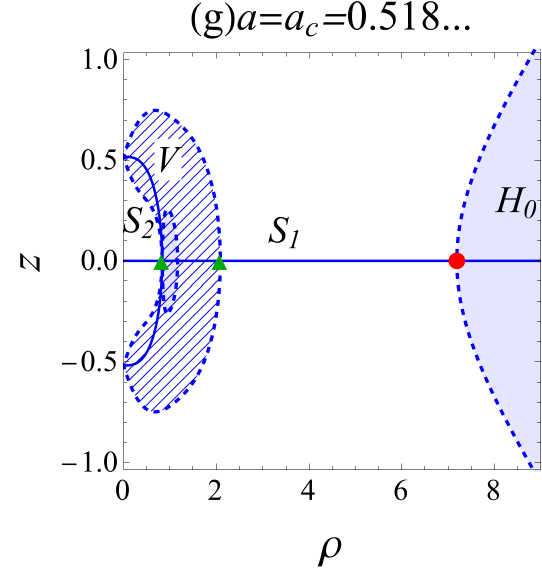}
    \end{minipage}
  \begin{minipage}[t]{0.3\columnwidth}
    \centering
       \includegraphics[width = 4cm]{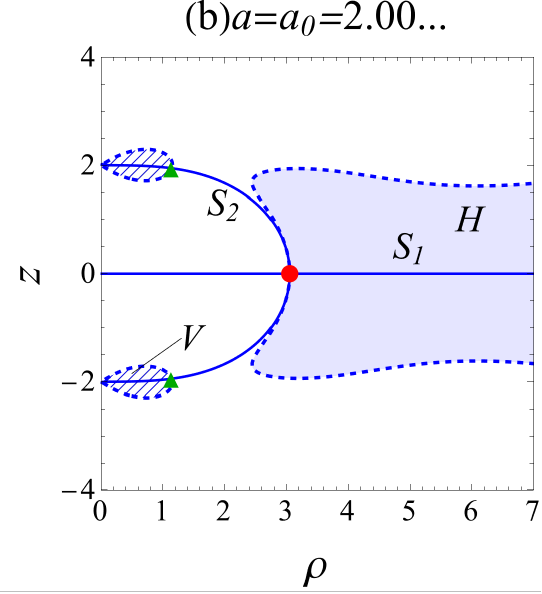}
 \includegraphics[width = 4cm]{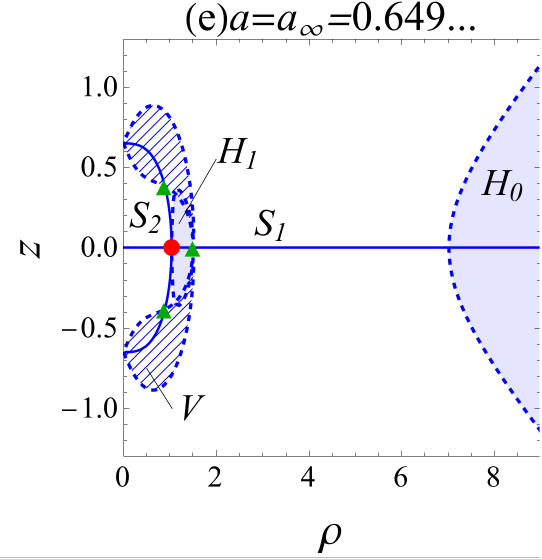}
   \includegraphics[width = 4cm]{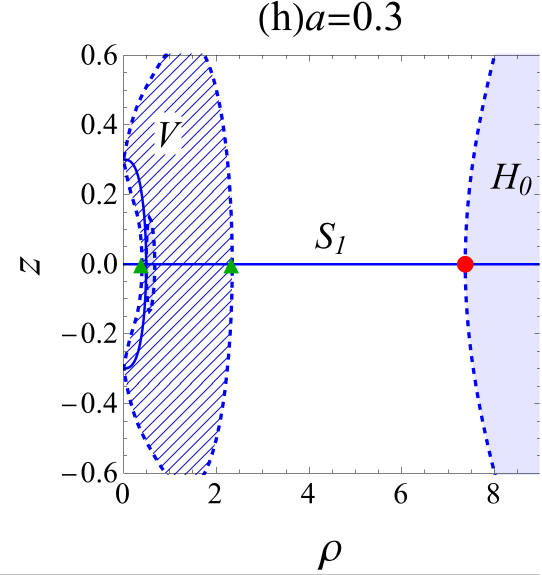}
  \end{minipage}
   \begin{minipage}[t]{0.3\columnwidth}
    \centering
       \includegraphics[width = 4cm]{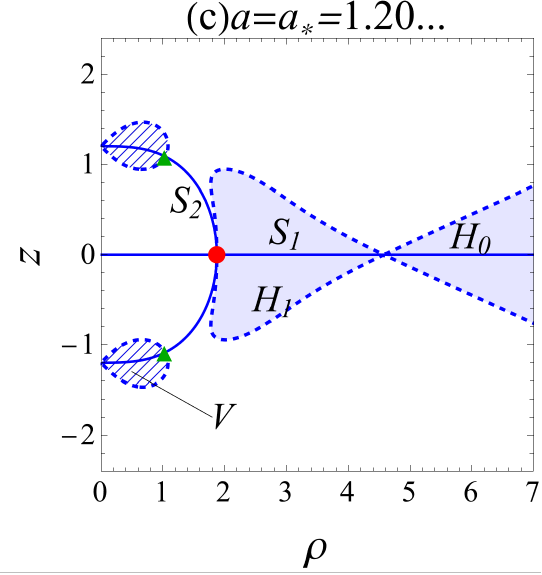}
   \includegraphics[width = 4cm]{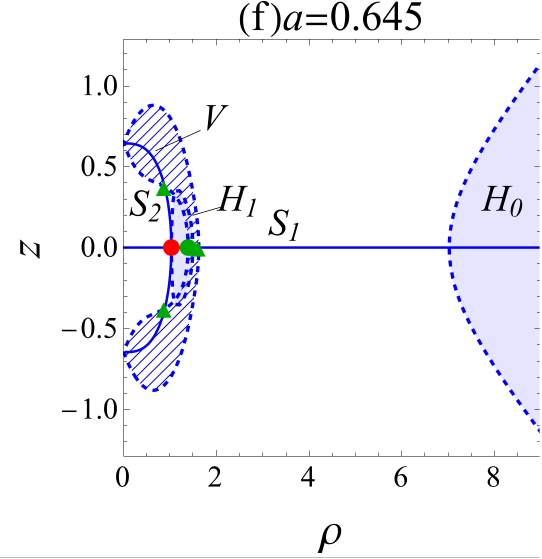}
 \end{minipage}

\caption{Circular orbits for massive and massless particles with $j<0$ and $J_1=J_2=0.499$.
Each panel differs in the separation $2a$ between two black holes located at $(\rho,z)=(0,\pm a)$. 
The sequence of stationary points $S$  comprises two curves, $S_1$ and $S_2$, represented by solid lines. 
Here, $S_1$ corresponds to the line $z=0$, while $S_2$ represents the arc connecting the two black holes. 
The red circle points represent ISCOs, while the green triangle and circle points denote unstable and stable circular orbits, respectively. 
Additionally, the red-colored region denotes the stability region $H$ and the hatched region denote the forbidden region $V$.
\label{j1+j20499-}}
\end{figure}

 For $0 < J_1 = J_2 < 0.5$ with $j < 0$, in general, the transition of orbits is qualitatively the same as in the case of $J_1=J_2=0.01$ with $j > 0$. However, the orbits appear at a greater distance from the black holes, as the counter-rotation of the particle intensifies the centrifugal effect. Here, we present two typical examples with $J_1=J_2=0.01$ and $J_1=J_2=0.499$ in Figs.~\ref{j1+j20.01-} and \ref{j1+j20499-}, respectively, which are qualitatively similar but with significantly different orbit positions.

\vspace{0.6cm}

\subsubsection{General results for $J_1=J_2$}

\begin{figure}[h]
\centering
\includegraphics[width=6cm]{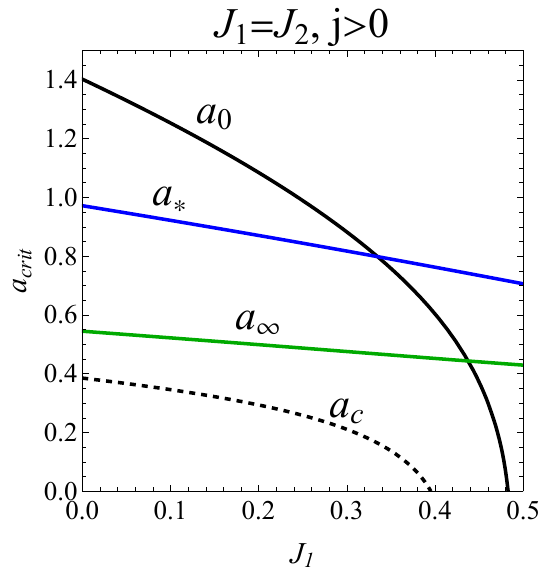}
\hspace{5mm}
\includegraphics[width=6cm]{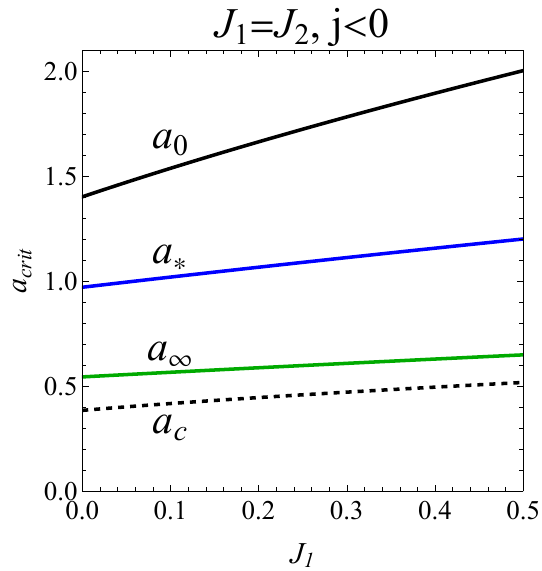}
\caption{Critical values of the black hole separation $a$ for $J_1=J_2$ with $j>0$ (the left panel) and $j<0$ (the right panel). In the left panel, one can find $a_c=0$ at $J_1=J_2=0.395\ldots$ and $a_0=0$ at $J_1=J_2=0.483\ldots$.
\label{fig:jacritJ1J1pp}}
\end{figure}

In Fig.~\ref{fig:jacritJ1J1pp}, we illustrate how each critical value of the separation $a$ depends on $J_1=J_2$ for each case of $j>0$ and $j<0$. For $j>0$, as shown in the left panel of Fig.~\ref{fig:jacritJ1J1pp}, the critical values monotonically decrease as the black holes' spin angular momenta becomes larger. 
Remarkably, the transition of ISCOs differs depending on the value of $J_1=J_2$, which can be classified into the following three patterns:

\begin{enumerate}
\item For $0\leq J_1=J_2 \leq 0.395...$, the ISCOs appear at
\begin{itemize}
\item[(a)] $a>a_0$ : the intersections of $S_2$ and the boundary of $H$ or $H_1$,
\item[(b)]  $a_c<a<a_0$ :  the intersection of $S_1$, $S_2$ and the boundary of $H$  or $H_1$,
\item[(c)] $a<a_c$:  the intersection of $S_1$ and the boundary of $H_0$.
\end{itemize}
\item For $0.395...<J_1=J_2<0.483...$, the ISCOs appear at
\begin{itemize}
\item[(a)] $a>a_0$ :  the intersections of $S_2$ and the boundary of $H_1$,
\item[(b)]  $0<a<a_0$ : the intersection of $S_1$, $S_2$ and the boundary of $H_1$,
\end{itemize}
\item For $0.483...<J_1=J_2<0.5$, the ISCOs appear at
\begin{itemize}
\item[(a)] $a>0$ :  the intersections of $S_2$ and the boundary of $H$  or $H_1$.
\end{itemize}
\end{enumerate}
In particular, the ISCOs stay on $S_2$ for arbitrary small $a$ if the spin angular momenta is sufficiently large $J_1=J_2>0.395\ldots$.

\medskip

In contrast, for $j<0$, as shown in the right panel of Fig.~\ref{fig:jacritJ1J1pp}, every critical value monotonically increases as the spin angular momenta become larger, and no significant change occurs in the transition of ISCOs.

\subsection{Black holes spinning in opposite directions}\label{subsec:oppositespin}

Now, let us consider the circular orbits around black holes spinning in opposite directions $(J_1=-J_2)$. Without loss of generality, we can assume $J_1=-J_2>0$. In this case, the circular orbits appear asymmetrically about $z=0$ but remain invariant under transformations $j\to -j$ and $z\to -z$. Therefore, we focus only on the $j>0$ case. Below, we present three typical cases: $J_1=-J_2=0.01$, $0.4$, and $0.499$, which represent all three possible scenarios.
Generally, the curve $S$ consists of two solid curves, $S_1$ and $S_2$. Here, $S_1$ is a curve starting from the black hole at $z=-a$ and approaching $z=0$ as $\rho \to \infty$, while $S_2$ is an arc connecting the origin $(\rho,z)=(0,0)$ and the black hole at $z=a$, as depicted in Fig.~\ref{j1-j2001}. The forbidden region $V$ and the stable region $H$ are represented by the hatched region and the red-colored region, respectively. Depending on the parameter, $H$ can split into two or three isolated components.
In the former case, we denote the part including infinity as $H_0$ and the other part as $H_1$. In the latter case, we denote the part including infinity as $H_0$, the part near $S_1$ as $H_1'$, and the part near $S_2$ as $H_2'$, as illustrated in Fig.~\ref{j1mj203}(g).
The ISCOs for massive particles are denoted by red circular points, and unstable and stable circular orbits for massless particles are represented by green triangles and circular points, respectively.

\subsubsection{$J_1=-J_2=0.01$}\label{sec:isco-j1-j2001}

\begin{figure}[h]
     \begin{minipage}[t]{0.3\columnwidth}
    \centering
\includegraphics[width = 4cm]{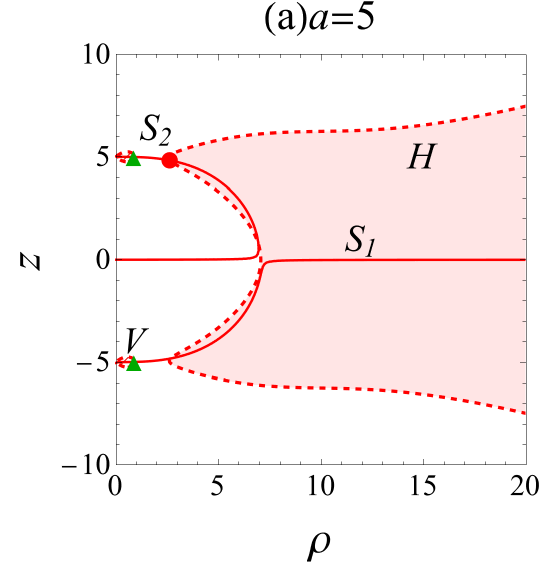}
   \includegraphics[width = 4cm]{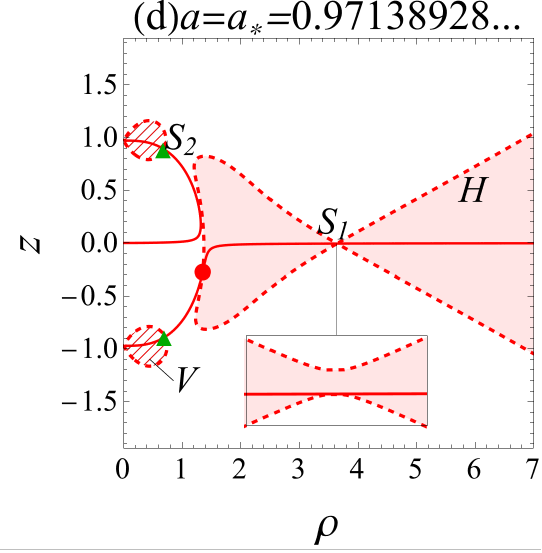}
    \includegraphics[width = 4cm]{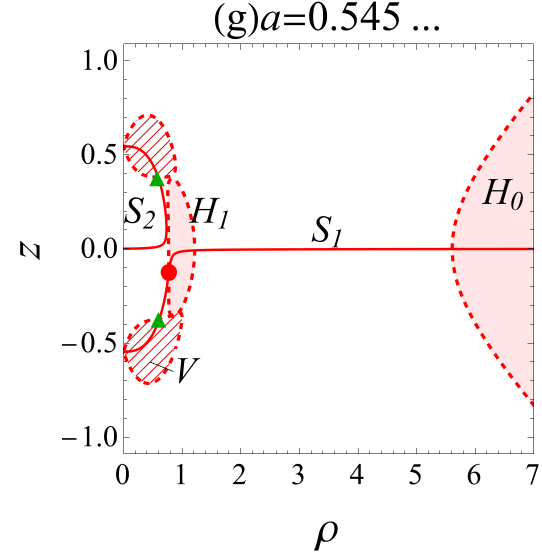}
   \includegraphics[width = 4cm]{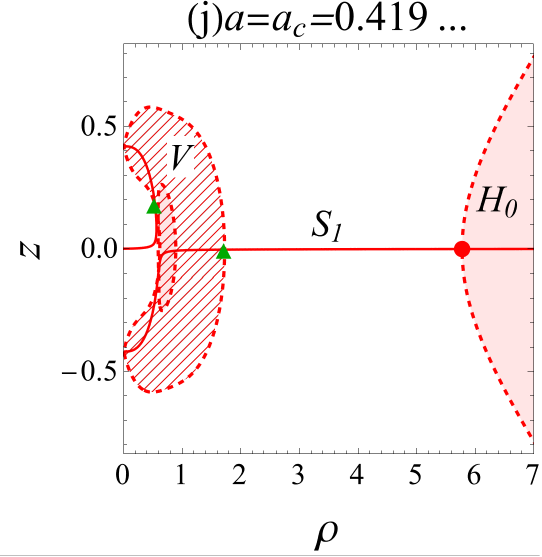}
 \end{minipage}
  \begin{minipage}[t]{0.3\columnwidth}
    \centering
   \includegraphics[width = 4cm]{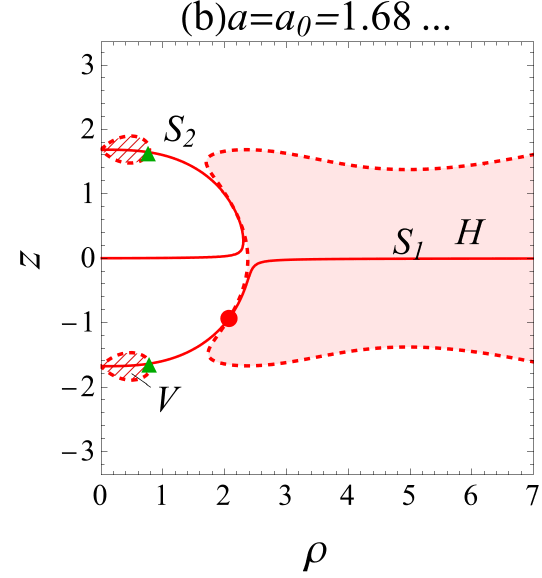}
      \includegraphics[width = 4cm]{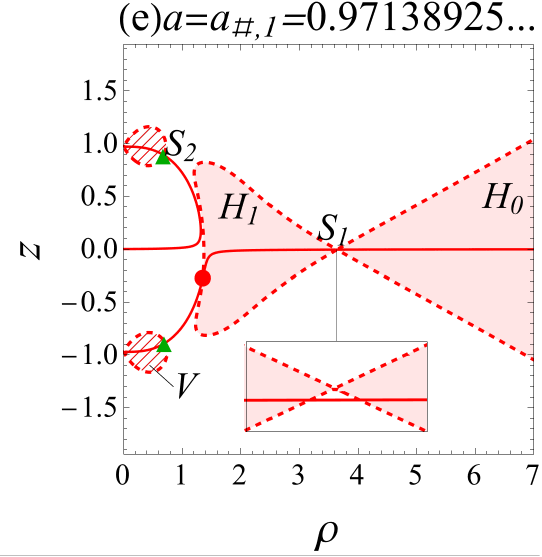}
 \includegraphics[width = 4cm]{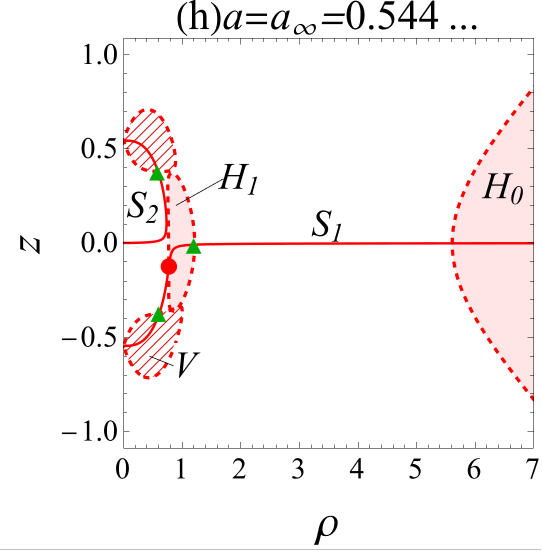}
   \includegraphics[width = 4cm]{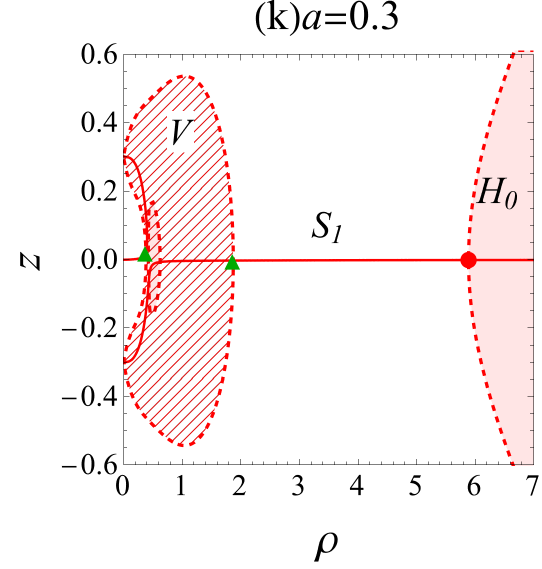}
\end{minipage}
   \begin{minipage}[t]{0.3\columnwidth}
    \centering
    \includegraphics[width = 4cm]{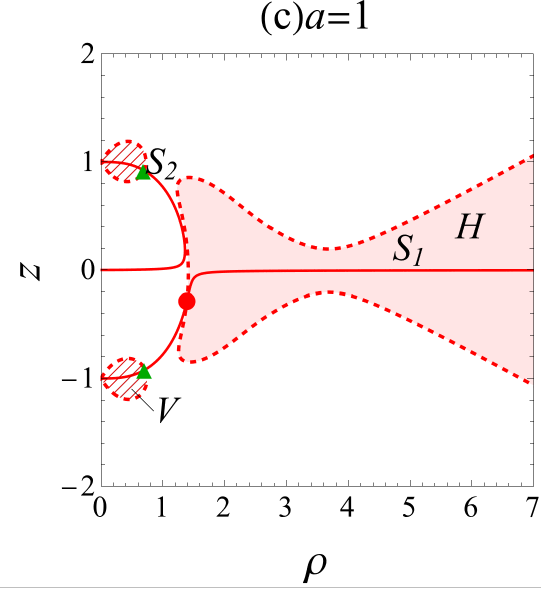}
       \includegraphics[width = 4cm]{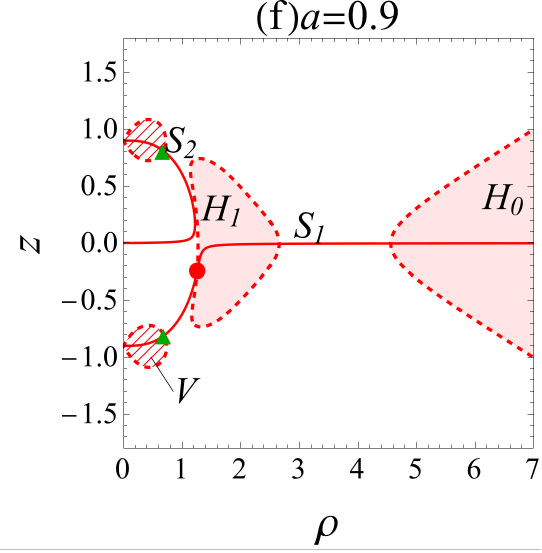}
\includegraphics[width = 4cm]{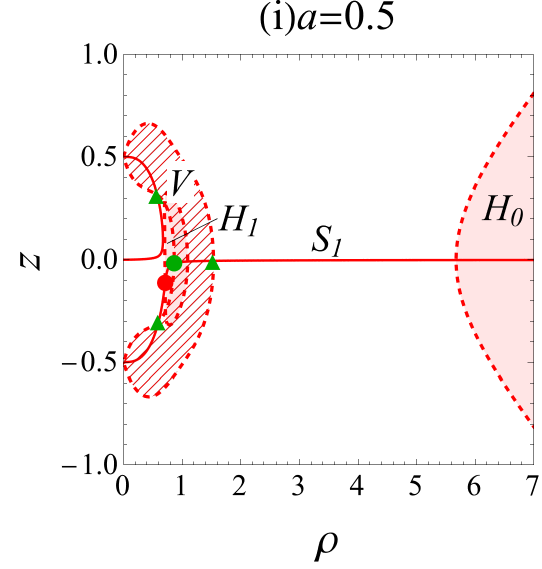}

 \end{minipage}
   \caption{Circular orbits for massive and massless particles with $j>0$ and $J_1=-J_2=0.01$.
 Each panel differs in the separation $2a$ between two black holes located at $(\rho,z)=(0,\pm a)$. 
The sequence of stationary points $S$ comprises two curves, $S_1$ and $S_2$, depicted by solid lines. 
Here, $S_1$ corresponds to the curve that starts from $(\rho,z)=(0,-a)$ and extends to $\rho\to\infty$ approaching $z=0$, while $S_2$ represents an arc connecting $(\rho,z)=(0,a)$ and the origin $(\rho,z)=(0,0)$. 
The red circle points represent the ISCOs, while the green triangle and circle points denote unstable and stable circular orbits, respectively. 
Additionally, the red-colored region denotes the stability region $H=H_0\cup H_1$, where $H_0$ includes infinity and $H_1$ does not,  and the hatched region denote the forbidden region $V$.
\label{j1-j2001}}
\end{figure}

First, we consider circular orbits for the $J_1=-J_2=0.01$ case, as presented in Fig.~\ref{j1-j2001}.
For $a>a_0=1.68...$, certain portions of $S_1$ and $S_2$ are included in $H$, allowing stable circular orbits for massive particles to exist on both $S_1$ and $S_2$. Since the smallest radius of stable circular orbits on $S_2$ is smaller than that on $S_1$, the ISCO is located at the intersection of $S_2$ and the boundary of $H$, as shown in Fig.~\ref{j1-j2001}(a). Unstable circular orbits for massless particles exist on both $S_1$ and $S_2$, where each intersects with the boundary of $V$ near each black hole.
At $a=a_0$, the boundary of $H$ becomes tangent to $S_2$, leading to the transfer of the ISCO to a point on $S_1$, as depicted in Fig.~\ref{j1-j2001}(b). This transition causes a discontinuous jump in the ISCO radius, contrasting with the continuous transition at $a=a_0$ in the binary with $J_1=J_2$ as seen in the previous section.
For $a_*=0.97138928...<a<a_0$, $H$ only intersects with $S_1$, and thus the ISCO exists solely on $S_1$ (Fig.~\ref{j1-j2001}(c)).
At $a=a_*$, the boundary of $H$ becomes tangent to $S_1$, forming a convex upward shape (Fig.~\ref{j1-j2001}(d)), leading to the division of the existence region of stable circular orbits for massive particles on $S_1$ into two parts for $a \leq a_*$. Slightly below $a_*$, at $a=a_{\#,1}=0.971380925...$, $H$ is also divided into two parts, where the isolated region closer to the black hole is denoted as $H_1$ and the part containing infinity is denoted as $H_0$ (Fig.~\ref{j1-j2001}(e)). Note that $a_*$ and $a_{\#,1}$ coincide for $J_1=J_2$, as observed in the previous section.

As shown in Fig.~\ref{j1-j2001}(h), at $a=a_\infty=0.544...$, another part of $V$ appears as a point at the outer intersection of $S_1$ and the boundary of $H_1$, leading to the emergence of a new unstable circular orbit for massless particles at this point. However, the ISCO remains at the intersection of $S_1$ and the boundary of $H_1$.
For $a_c=0.418<a<a_\infty$, a stable circular orbit for massless particles exists at the intersection of $S_1$ and the boundary of $V$ inside $H_1$, in addition to the other three unstable orbits (Fig.~\ref{j1-j2001}(i)).
At $a = a_c$, as depicted in Fig.~\ref{j1-j2001}(j), the boundary of $V$ becomes tangent to $S_1$ and $S_1$ is fully covered by $V$, causing the ISCO to jump to the intersection of $S_1$ and the boundary of $H_0$. Consequently, the stable circular orbit and an unstable circular orbit of massless particles on $S_1$ disappear, similar to the binary with $J_1=J_2$.
For $a< a_c$, as shown in Fig.~\ref{j1-j2001}(k), the ISCO remains at the intersection of $S_1$ and the boundary of $H_0$, while massless particles have two unstable circular orbits: one at the intersection of $S_2$ and the boundary of $V$, and the other at the intersection of $S_1$ and the boundary of $V$.

\subsubsection{$J_1=-J_2=0.3$}

\begin{figure}[h]
     \begin{minipage}[t]{0.3\columnwidth}
    \centering
   \includegraphics[width = 4cm]{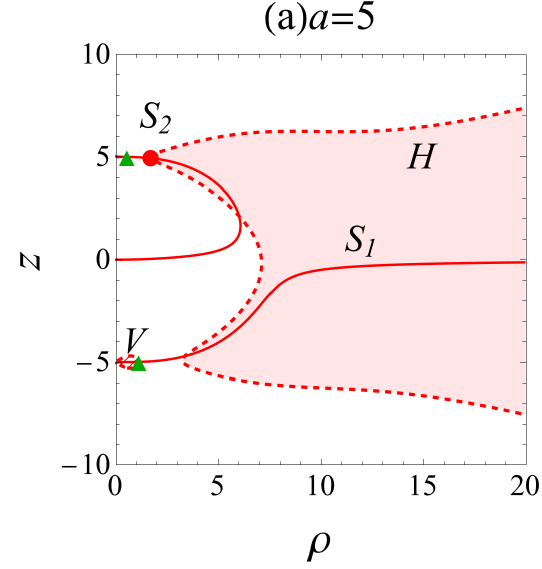}
    \includegraphics[width = 4cm]{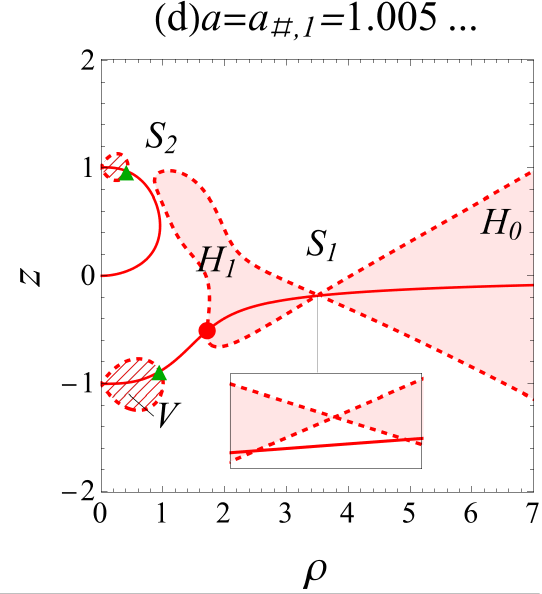}
\includegraphics[width = 4cm]{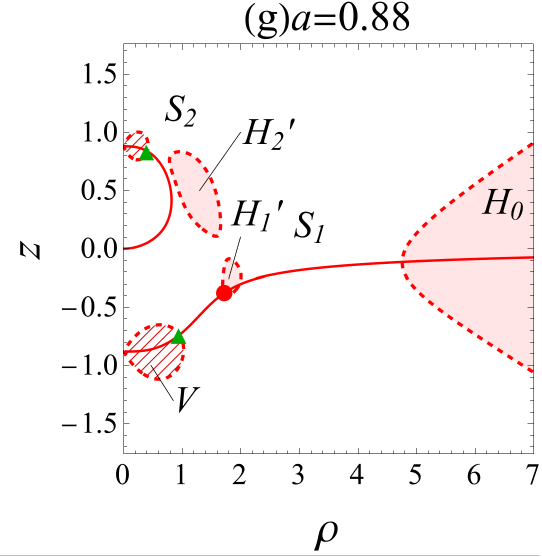}
\includegraphics[width = 4cm]{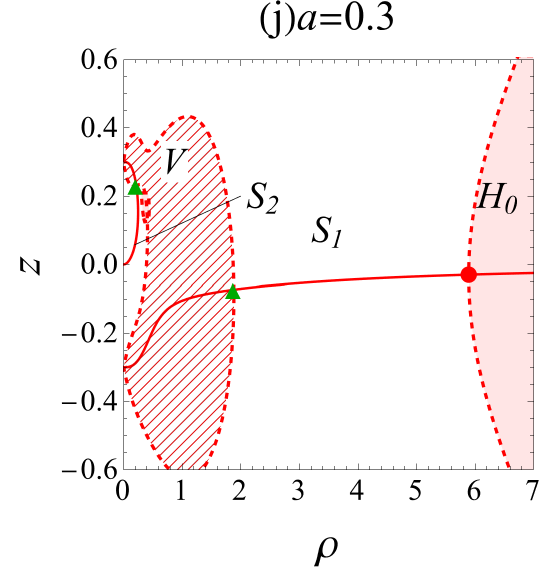}
  \end{minipage}
  \begin{minipage}[t]{0.3\columnwidth}
    \centering
    \includegraphics[width = 4cm]{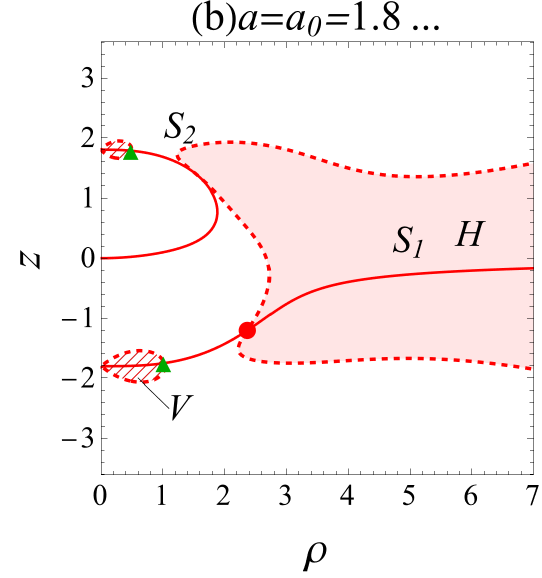}
     \includegraphics[width = 4cm]{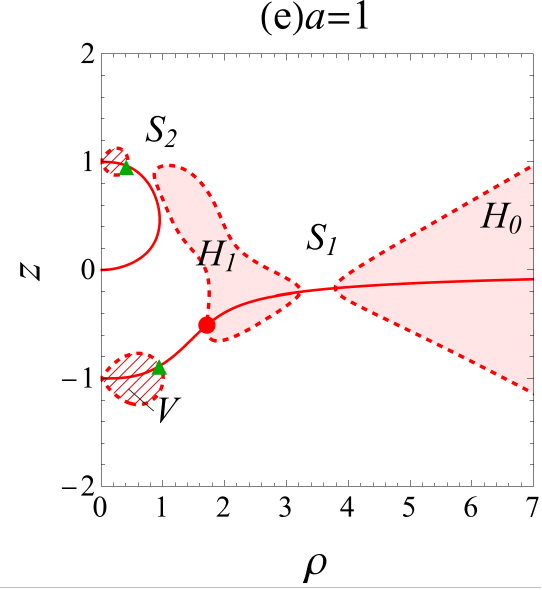}
\includegraphics[width = 4cm]{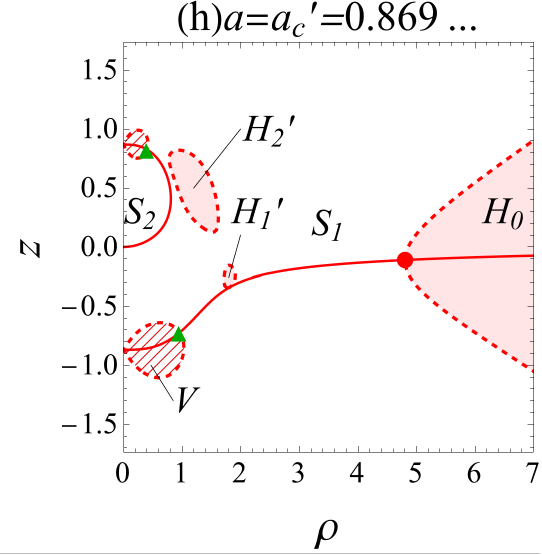}

   \end{minipage}
   \begin{minipage}[t]{0.3\columnwidth}
    \centering
 \includegraphics[width = 4cm]{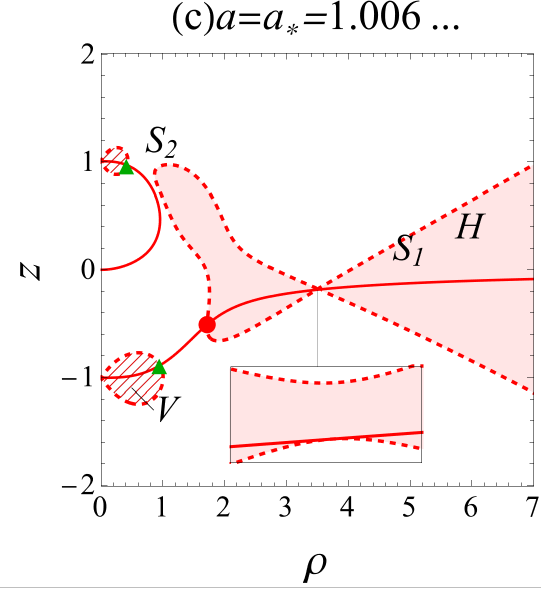}
 \includegraphics[width = 4cm]{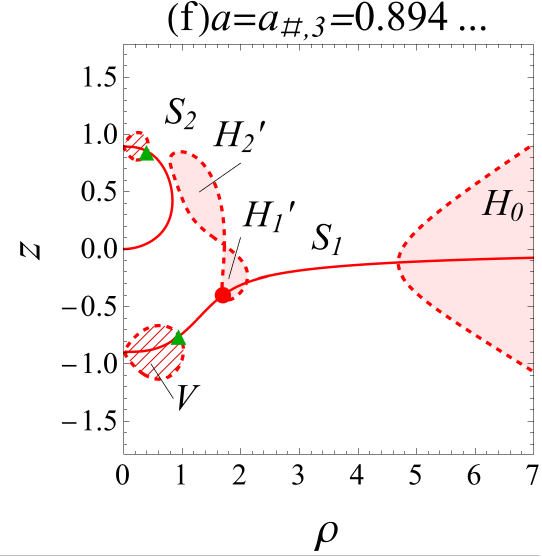}
\includegraphics[width = 4cm]{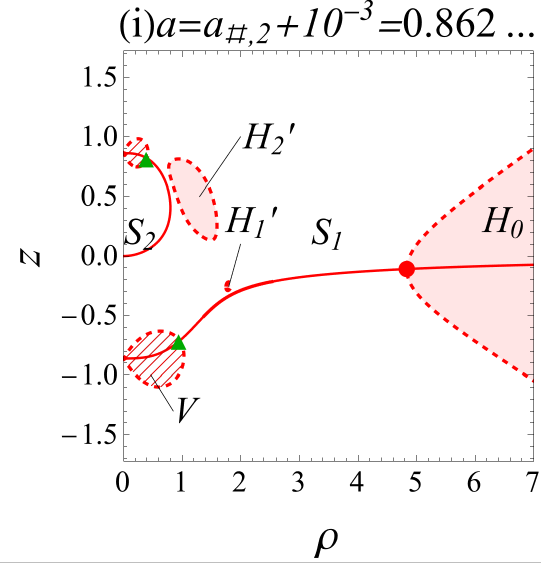}
  \end{minipage}
    \caption{
    Circular orbits for massive and massless particles with $j>0$ and $J_1=-J_2=0.3$. 
     Each panel differs in the separation $2a$ between two black holes located at $(\rho,z)=(0,\pm a)$. 
The sequence of stationary points $S$ comprises two curves, $S_1$ and $S_2$, depicted by solid lines. 
Here, $S_1$ corresponds to the curve that starts from $(\rho,z)=(0,-a)$ and extends to $\rho\to\infty$ approaching $z=0$, while $S_2$ represents an arc connecting $(\rho,z)=(0,a)$ and the origin $(\rho,z)=(0,0)$.  
The red circle points represent the ISCOs, while the green triangle and circle points denote unstable and stable circular orbits, respectively. 
Additionally, the red-colored region denotes the stability region $H=H_0\cup H_1$ or $H_0 \cup H_1' \cup H_2'$, where $H_0$ includes the infinity, $H_1'$ denotes the part near $S_1$ and $H_2'$ near $S_2$. The hatched region denote the forbidden region $V$. 
\label{j1mj203}}
\end{figure}

Next, we consider the case of $J_1=-J_2=0.3$.
As depicted in Fig.~\ref{j1mj203}(a)-(e), the transition of circular orbits follows the same pattern as in the $J_1=-J_2=0.01$ case but at different values of $a_0$, $a_*$, and $a_{\#,1}$.
In Fig.~\ref{j1mj203}(f), at $a=a_{\#,3}=0.894...$, $H_1$ further splits into two parts $H_1'$ and $H_2'$, with $H_1'$ closer to $S_1$ and $H_2'$ closer to $S_2$. However, since $H_1'$ intersects with $S_1$ instead of $H_1$, this topological change does not affect the ISCO and the existence region of stable circular orbits on $S_1$.
Therefore, for $a_{c}'=0.869...<a\leq a_*$, the existence region of stable circular orbits on $S_1$ consists of two parts, and the ISCO lies at the boundary of the part not connected to infinity (Figs.~\ref{j1mj203}(c)-(g)).
At $a=a_c'$, depicted in Fig.~\ref{j1mj203}(h), the boundary of $H_1'$ tangentially touches $S_1$, and the stable circular orbits exist only within $H_0$. This transition causes a jump of the ISCO from the boundary $H_1'$ to the boundary of $H_0$, similar to the transition at $a=a_c$ for the $J_1=-J_2=0.01$ and $J_1=J_2$ cases, but occurring through a different process due to the latter occurring when $H_1$ is fully covered by $V$.
For $a<a_{c}'$, $H_1'$ and $H_2'$ never intersect with $S_1$ or $S_2$. Consequently, the stable circular orbits appear only within $H_0$ on $S_1$, and the ISCO remains on the boundary of $H_0$.
At $a=a_{\#,2}=0.861...$, as shown in Fig.~\ref{j1mj203}(i), $H_1'$ shrinks to a point and disappears. Therefore, $H$ has only two parts, $H_0$ and $H_1$, for $a<a_{\#,2}$, where $H_1$ does not intersect $S_1$ or $S_2$ and is finally included in $V$ (Fig.~\ref{j1mj203}(j)).
Additionally, the massless particles admit only two unstable circular orbits around each black hole for any $a$, with one on $S_1$ and another on $S_2$.

\subsubsection{$J_1=-J_2=0.499$}

\begin{figure}[h]
     \begin{minipage}[t]{0.3\columnwidth}
    \centering
 \includegraphics[width = 4cm]{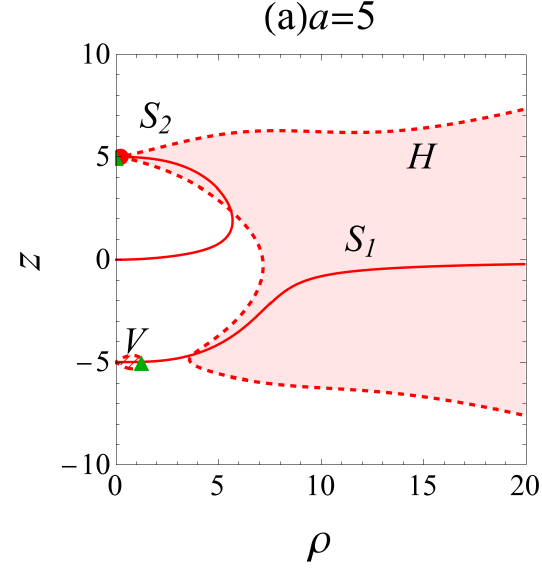}
    \includegraphics[width = 4cm]{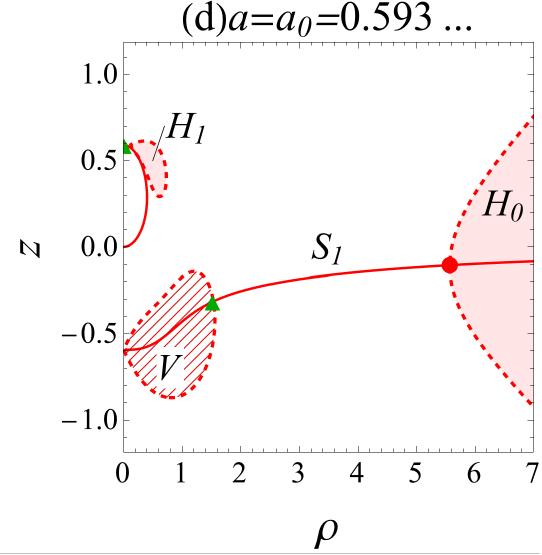}
  \end{minipage}
  \begin{minipage}[t]{0.3\columnwidth}
    \centering
 \includegraphics[width = 4cm]{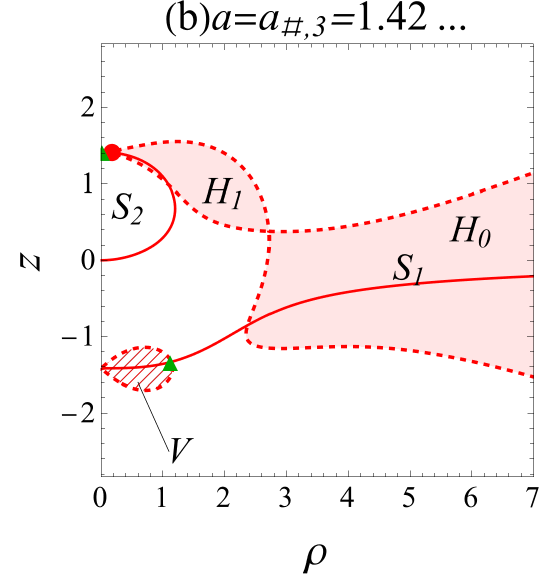}
   \includegraphics[width = 4cm]{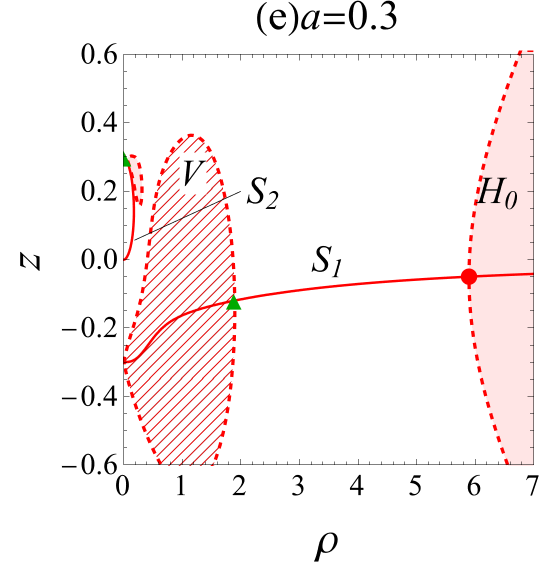}
  \end{minipage}
   \begin{minipage}[t]{0.3\columnwidth}
    \centering
\includegraphics[width = 4cm]{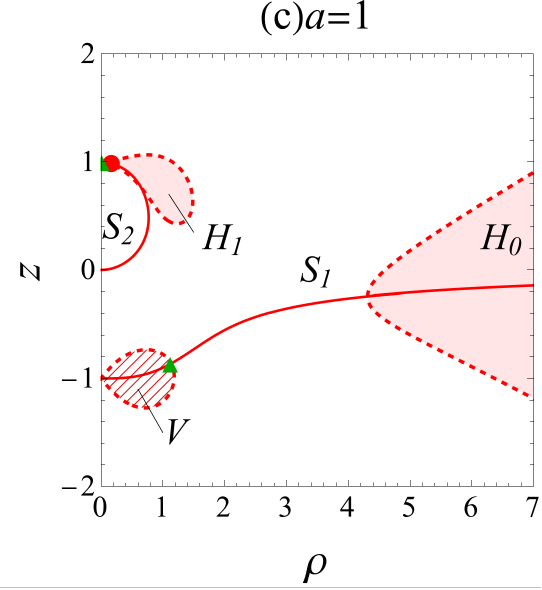}
  \end{minipage}

   \caption{Circular orbits for massive and massless particles with $j>0$ and $J_1=-J_2=0.499$.
     Each panel differs in the separation $2a$ between two black holes located at $(\rho,z)=(0,\pm a)$. 
The sequence $S$ of stationary points comprises two curves, $S_1$ and $S_2$, depicted by solid lines. 
Here, $S_1$ corresponds to the curve that starts from $(\rho,z)=(0,-a)$ and extends to $\rho\to\infty$ approaching $z=0$, while $S_2$ represents an arc connecting $(\rho,z)=(0,a)$ and the origin $(\rho,z)=(0,0)$.  
The red circle points represent the ISCOs, while the green triangle and circle points denote unstable and stable circular orbits, respectively. 
Additionally, the red-colored region denotes the stability region $H=H_0\cup H_1$, where $H_0$ includes infinity and $H_1$ does not,  and the hatched region denote the forbidden region $V$.
   \label{j1-j20499}}
\end{figure}

Finally, let us consider the case of $J_1=-J_2=0.499$, which differs drastically from the previous two cases.
From Fig.~\ref{j1-j20499}, we can observe the following:
For sufficiently large $a$, as shown in Fig.~\ref{j1-j20499}(a), both $S_1$ and $S_2$ are included in the stable region $H$, and stable circular orbits exist on both curves. By comparing the minimum radii of stable circular orbits on $S_1$ and $S_2$, we find that the ISCO lies on $S_2$.
At $a=a_{\#, 3}=1.42...$, depicted in Fig.~\ref{j1-j20499}(b), $H$ splits into two parts, $H_1$ and $H_0$, where $H_0$ extends to infinity and $H_1$ is closer to $S_2$. However, this topology change occurring away from $S_1$ or $S_2$ does not affect the topology of stable circular orbits.
For $a_0=0.593...<a<a_{\#,3}$, $H_1$ contains a part of $S_2$ instead of $H$, as shown in Fig.~\ref{j1-j20499}(c), and hence the ISCO remains on $S_2$.
At $a=a_0$, the boundary of $H_1$ tangentially touches $S_2$, leading to the ISCO transitioning from a point on $S_2$ to a point on $S_1$ as depicted in Fig.~\ref{j1-j20499}(d). For $a\leq a_0$, the ISCO stays at the intersection of $S_1$ and the boundary of $H_0$.
Meanwhile, similar to the $J_1=-J_2=0.3$ case, massless particles have an unstable circular orbit on $S_1$ and another on $S_2$ for any $a$.

\subsubsection{General results for $J_1=-J_2$}

\begin{figure}[h]
\includegraphics[width=7cm]{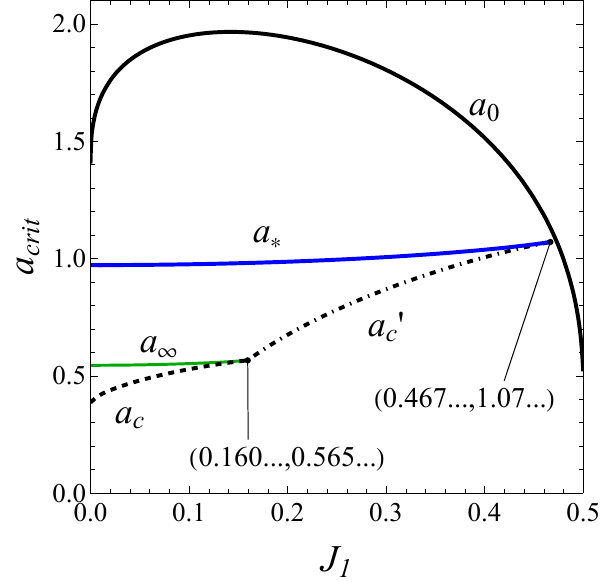}
\hspace{5mm}
\includegraphics[width=7cm]{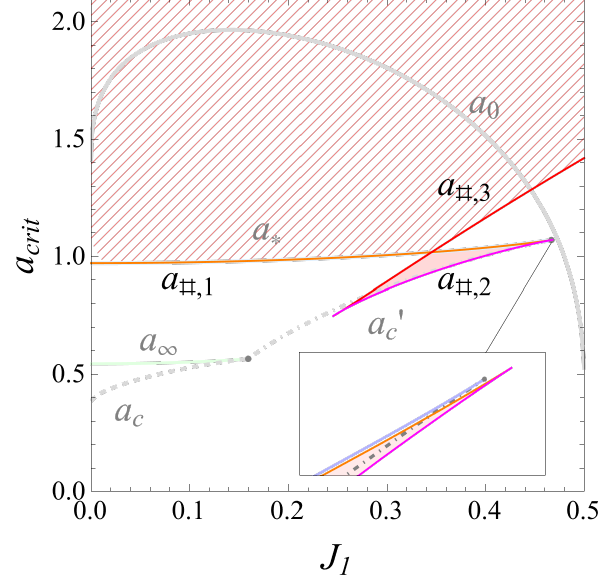}
\caption{
Critical values of the black hole separation $a$ for $J_1=-J_2$ responsible for the orbit transition (left panel) and topology change of the stable region $H$ (right panel). In the right panel, $H$ has only one connected part in the hatched region, two in the white region and three in the red region. $a_*$ and $a_{\#,1}$ seems to coincide but $a_*$ is always slightly greater than $a_{\#,1}$ in the close-up.
Similarly, a part of $a_{c}'$ and $a_{\#,2}$ seems quite close but $a_{c}'$ is always greater.
\label{fig:acrit-1}
}
\end{figure}

In general, the phase of circular orbits in the binary with $J_1=-J_2$ fits into one of the three patterns described earlier. In particular, from Fig.~\ref{fig:acrit-1}, we can observe that the transition of the ISCO occurs differently for the following ranges of $J_1=-J_2>0$.

\begin{figure}[h]
\begin{center}
\includegraphics[width=6cm]{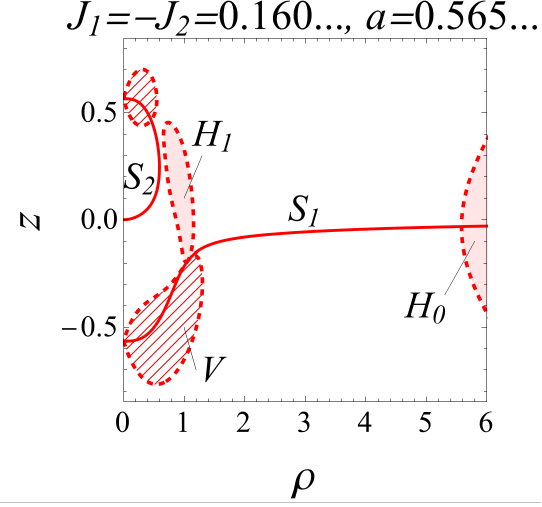}
\hspace{5mm}
\includegraphics[width=5.8cm]{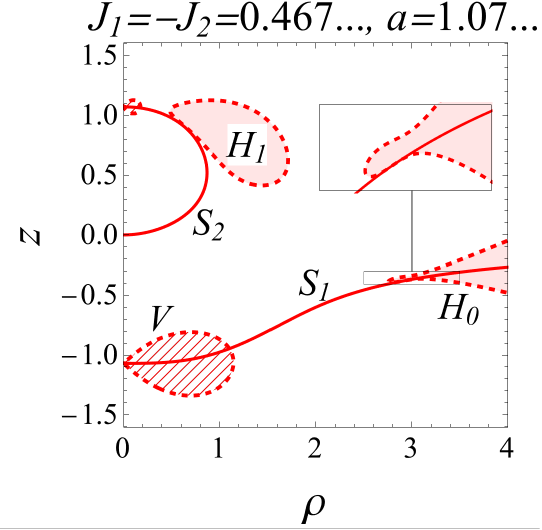}
\caption{Circular orbits for critical parameters $(J_1,a)=(0.160...,0.565...)$ and $(J_1,a)=(0.467...,1.07...)$. In the left panel, $S_1$, the boundary of $H_1$ and $V$ are all tangent to each at a point. In the right panel, $S_1$ and the boundary of $H_0$ are tangent
up to second derivatives.
 \label{fig:acrit-H}}
\end{center}
\end{figure}

\begin{enumerate}
\item For $0< J_1<0.160...$, the ISCO appears at
  \begin{itemize}
  \item[(a)] $a>a_0$ : the intersection of  $S_2$ and the boundary of $H$,
    \item[(b)] $a_c<a<a_0$ : the innermost intersection of $S_1$ and the boundary of $H$ or $H_1$,
        \item[(c)] $a<a_c$ : the intersection of  $S_1$ and the boundary of $H_0$,
  \end{itemize}
\item For $0.160<J_1<0.467...$, the ISCO appears at
  \begin{itemize}
  \item[(a)] $a>a_0$ : the intersection of  $S_2$ and the boundary of $H$,
    \item[(b)] $a_c'<a\leq a_0$ : the innermost intersection of  $S_1$ and the boundary of $H$ or $H_1$ or $H'_1$,
        \item[(c)] $a\leq a_c'$ : the intersection of  $S_1$ and the boundary of $H_0$,
  \end{itemize}
\item For $0.467<J_1<0.5$,  the ISCO appears at
  \begin{itemize}
  \item[(a)] $a>a_0$ : the intersection of  $S_2$ and the boundary of $H$ or $H_1$,
    \item[(b)] $a\leq a_0$ : the intersection of  $S_1$ and the boundary of $H_0$,
  \end{itemize}
\end{enumerate}
where each case corresponds to $J_1=0.01,0.3,0.499$.
The three critical curves $a_\infty$, $a_{c}$, and $a_{c}'$ intersect at $(J_1,a)=(0.160..., 0.565...)$, where $S_1$ intersects with the boundary of $H_1$ and $V$ at a single point (the left panel in Fig.~\ref{fig:acrit-H}).
Similarly, the two critical curves $a_{c}'$ and $a_*$ intersect at $(J_1,a)=(0.467...,1.07...)$, where $S_1$ becomes tangent to the boundary of $H_0$ up to second derivatives (the right panel in Fig.~\ref{fig:acrit-H}).
It is worth noting that $a_{c}'$ and $a_*$ do not extend beyond $J_1=0.467\ldots$ without reaching $a_0$.

\medskip

In the right panel of Fig.~\ref{fig:acrit-1}, we present the critical values relevant to the topology change of $H$. Although these values are not directly involved in the transition of orbits, they are useful for understanding the appearance of stable circular orbits.
The curves $a_{\#,1}$ and $a_{\#,3}$ correspond to the critical values where $H$ splits into two parts. The former corresponds to situations depicted in Fig.~\ref{j1-j2001}(e), Fig.~\ref{j1mj203}(d), and similar cases. The latter corresponds to situations depicted in Fig.~\ref{j1mj203}(f), Fig.~\ref{j1-j20499}(b), and similar scenarios.
The curve $a_{\#,2}$ represents the critical value where an isolated component of $H$ shrinks to a point and disappears, as shown in Fig.~\ref{j1mj203}(i). Interestingly, $a_{\#,1}$ and $a_{\#,2}$ meet at $(J_1,a) = (0.245...,0.748...)$, while $a_{\#,2}$ and $a_{\#,3}$ meet at $(J_1,a)=(0.468...,1.07...)$.

\section{Summary and discussion}\label{sec:summary}

In this paper, we have investigated the motion of particles around two spinning charged black holes by using the Teo-Wan solution. We have shown that the motion of both massive and massless particles around this binary black hole system can be reduced to a two-dimensional potential problem.
We have made assumptions that the masses of the two black holes are equal $M_1=M_2=m$, the directions of the spin angular momenta are aligned with the line connecting the two black hole centers, and the magnitudes of the spin angular momenta are equal.  
Subsequently, we divided the analysis into cases where the two black holes are spinning in the same direction and in opposite directions, and examined the stability of particle's circular motion. Since the upper limit of the spin angular momenta of the black holes is $0.5m^2$, which corresponds to a singular solution, we investigated how the range of existence of particle's circular motion changes as we varied the spin angular momenta $|J_1|,|J_2|$ of the black holes from $0$ to $0.499 m^2$, with the distance between the black holes as a parameter.

\medskip

First, for binary black holes rotating in the same direction, the shape of the curve $S$, which represents a set of stationary points of the effective potential (which denotes a set of circular orbits in the allowed region $U\le 0$), could be described as a straight line at $z=0$ and a semicircular curve connecting the black holes, similar to the case when there is no rotation of the black holes.
Particles rotating in the same direction as the black holes exhibit smaller radii for both the ISCO and the circular orbits of massless particles compared to when there is no rotation. Additionally, as the magnitude of the black holes' spin angular momentum increases, these radii become smaller. On the other hand, particles rotating in the opposite direction to the black holes exhibit larger radii for both the ISCO and the circular orbits of massless particles compared to when there is no rotation. Moreover, as the magnitude of the black holes' spin angular momentum increases, these radii become larger.
For particles rotating in the same direction as the black holes, we also found that the transition of the ISCO occurs differently in the  three ranges of angular momenta, $0<J_1/m^2=J_2/m^2< 0.395\ldots$, $0.395\ldots<J_1/m^2=J_2/m^2< 0.483\ldots$ and $0.483\ldots<J_1/m^2=J_2/m^2<0.5$, while for particles rotating in the opposite direction to the black holes, the transitoin occurs in the same
manner for $0<J_1/m^2=J_2/m^2<0.5$.

\medskip

Next, for binary black holes rotating in opposite directions, the curve $S$, which is a set of stationary points of the effective potential, consists of two curves. One curve connects points where particles exist with black holes rotating in the same direction to the origin, while the other curve asymptotically approaches $z=0$ from points where black holes rotate in opposite directions.
As a common feature independent of the magnitude of angular momentum, it was found that when the distance 
$2a$ between the black holes is large, the ISCOs exist on the arched curve, and when the distance is small, they approach 
$z=0$ along the curve. Furthermore, it was found that reducing the distance between the black holes leads to a discontinuous change in the radius of ISCO, causing it to increase. 
Additionally, for the non-rotating black holes, it was found that at certain values of $a$, massless particles have only up to four circular orbits. However, when the spins of the black holes are both small in magnitude ($0<J_1/m^2=-J_2/m^2<0.160\ldots$),  it was found that at certain values of $a$, 
there exist, at a maximum, a total of eight simultaneous circular orbits for massless particles. Among these, as discussed in Sec.~\ref{sec:isco-j1-j2001}, four correspond to particles with $j > 0$, while the remaining four correspond to particles with $j < 0$.
We also   found that the transition of the ISCO occurs differently in the three ranges of angular momenta, $0<J_1/m^2=-J_2/m^2< 0.160\ldots$, $0.160\ldots<J_1/m^2=-J_2/m^2< 0.467\ldots$ and $0.467\ldots<J_1/m^2=-J_2/m^2<0.5$.

\medskip

These phenomena are considered to be caused by the dragging effect of the spinning black holes.
The space around a spinning black hole is dragged by the black hole, causing particles that are stationary at the location near the black hole to appear rotating when viewed from afar. 
In other words, around a spinning black hole, particles rotate in the same direction as the black hole.
As a result of this effect, particles rotating in the same direction as the black hole behave as if they are rotating faster. Therefore, it is expected that they can rotate at positions closer to the center. 
Conversely, particles rotating in the opposite direction to the binary black holes appear to rotate slower, hence it is expected that their rotating radius becomes larger.
In this paper, for simplicity, we assumed that black holes have equal mass, and equal spin angular momenta. 
Investigating the case with different masses and spin angular momenta is a topic for our future research.

\acknowledgments

RS was supported by JSPS KAKENHI Grant Number JP18K13541. 
ST was supported by JSPS KAKENHI Grant Number 21K03560.




\end{document}